\documentclass[preprint,showpacs,preprintnumbers,amsmath,amssymb,nofootinbib]{revtex4}
\usepackage{latexsym}
\usepackage{graphicx}
\usepackage{dcolumn}
\usepackage{bm}
\usepackage{xcolor}
\usepackage{dcolumn}   
\usepackage{multirow}	 
\usepackage{rotating}
\begin{document}
\title{Diffusion-reaction model for positron trapping and annihilation at spherical extended  defects
and in precipitate$-$matrix composites}

\author{Roland W\"{u}rschum, Laura Resch, and Gregor Klinser}
\affiliation{ Institute of Materials Physics, Graz University of Technology, Petersgasse 16, A-8010 Graz, Austria}

\email{wuerschum@tugraz.at}

\date{Published 26 June 2018} Phys. Rev. B {\bf 97} (2018) 224108;  DOI: 10.1103/PhysRevB97.224108

\begin{abstract}
The exact solution of a diffusion$-$reaction model for the
trapping and annihilation of positrons in small extended spherical defects (clusters, voids, small precipitates)  
with competitive rate-limited trapping in vacancy-type point defects
is presented. Closed-form expressions are obtained for the mean positron lifetime and
for the intensities of the two positron lifetime components associated with
trapping at defects. The exact solutions can be conveniently applied for the analysis of experimental data
and  allow an assessment in how far the usual approach, which takes  diffusion limitation into account 
by means of effective diffusion-trapping rates, is appropriate. 
The model is further extended for application to larger precipitates where
diffusion- and reaction limited trapping is not only considered for the trapping
from the matrix into the  precipitate$-$matrix interface but also for 
the trapping from inside the precipitates into the interfaces.
This makes the model applicable to all type
of composite structures where spherical objects are embedded in a matrix irrespective of their size and  their number density.
\end{abstract}


\maketitle

\section{Introduction}
The versatile technique of positron annihilation makes use of the fact that positrons ($e^+$) are trapped at  free 
volume-type defects which allows their detection by a specific variation of the positron-electron annihilation characteristics \cite{Hautojaervi79, Brossmann12, Krause-Rehberg99, Puska94}. 
Whereas the kinetics of $e^+$ trapping at vacancy-type point defects can be well described by 
rate theory (so-called simple trapping model), it is well known that for trapping at extended defects like
grain boundaries, interfaces, voids, clusters, or precipitates, diffusion limitation of the trapping process may be an issue. 
Diffusion-limited positron trapping at interfaces and grain boundaries has been quantatively modeled by several groups,
ranging from entirely diffusion-controlled trapping \cite{Brandt72}, diffusion-reaction controlled trapping including detrapping
\cite{Dupasquier93, Wuerschum96, Koegel96, dryzek1999}, up to diffusion-reaction controlled trapping at grain boundaries and competetive transition-limited trapping at point defects in crystals 
\cite{Cizek02, Oberdorfer09, dryzek1998, Koegel96}.

Compared to grain boundaries, diffusion-limited $e^+$ trapping at voids and clusters has not been studied in such detail 
despite the undoubted relevance of positron annihilation for studying this important class of defects \cite{Nieminen79a, Bentzon90, Eldrup03}.
One approach to deal with diffusion-limited trapping is based on effective diffusion-trapping rates 
which then allow an implementation in standard rate theory  (e.g., \cite{Bentzon90}).
Diffusion-limited trapping at point-like defects was studied by Dryzek \cite{dryzek1998diffusion}
for the one-dimensional case. A full treatment of $e^+$ trapping and annihilation in voids
in the framework of diffusion-reaction theory was given by Nieminen et al. \cite{Nieminen79}.
This treatment of Nieminen et al. \cite{Nieminen79} is conceptionally analogous to the subsequent work of Dupasquier et al. 
\cite{Dupasquier93} for diffusion-limited $e^+$ trapping at grain boundaries, both of which lead to solutions exclusively in terms of infinite series.

Another treatment  of the diffusion-reaction problem of $e^+$ trapping at grain boundaries
was given by W\"urschum and Seeger \cite{Wuerschum96} which yields closed-form expressions
for the mean $e^+$ lifetime and the intensity of the annihilation component associated with 
the trapped state. This approach is applied in the present work to the diffusion-reaction problem 
of $e^+$ trapping and annihilation in spherical extended  defects (voids, clusters, precipitates).\footnote{For the sake of
simplicity, representatively for all kinds of  spherical extended defects (voids, clusters, or precipitates) 
the term voids is used in the following.} 
Following our earlier further work on grain boundaries 
\cite{Oberdorfer09}, now in addition competitive reaction rate-limiting trapping at point defects is taken into account.
The present  treatment yields closed-form expressions of the major $e^+$ annihilation parameters
for this application-relevant case of competitive
$e^{+}$ trapping in voids and point defects.
These closed-form expressions allow deeper insight in the physical details of
$e^+$ annihilation characteristics as well as an assessment of the so far often used  approach based on
effective diffusion-trapping rates. Above all, the results can be conveniently applied for
the analysis of experimental data.

In a further part, the model presented here and the previous model on positron trapping at grain boundaries
are merged in order to study precipitates embedded in matrix.  
Here, diffusion- and reaction limited trapping is considered for both the trapping
from the matrix into the  precipitate$-$matrix interface and for  
the trapping from inside the precipitates into the interfaces.

\section{The Model}
\begin{figure}[ht]
\includegraphics[width=8.5cm]{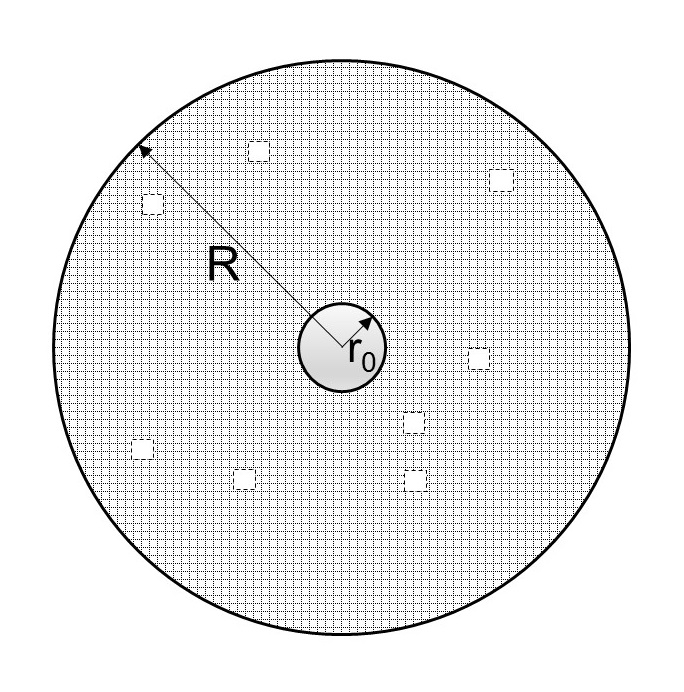}
\caption{Geometry of the diffusion-reaction model: spherical voids of radius $r_0$ are located in a lattice
with homogeneously distributed vacancy-type defects ($\Box$) in which reaction-controlled trapping occurs.
The outer radius $R$ of the diffusion sphere defines the void concentration.
}
\label{fig:1}
\end{figure}

The model describes positron ($e^+$) trapping and annihilation in  voids in the general case that both the $e^+$ diffusion and the transition reaction has to be taken into account (so called diffusion-reaction controlled trapping process).
In order to cover more complex cases, competitive transition-limited trapping at vacancy-type points defects is also considered
(see Fig. \ref{fig:1}).
This procedure follows our earlier study 
where concomitant positron trapping at grain boundaries and at point defects in crystallites has been considered \cite{Oberdorfer09}.

The behavior of the positrons is described by their bulk (free) lifetime $\tau_f$,  
by their lifetime ($\tau_t)$ in the  voids, 
by their lifetime ($\tau_v)$ in the vacancy-type point defects in the lattice (matrix),
and by their bulk diffusivity $D$.
Trapping at the point defects of the matrix is characterized by the specific $e^+$ trapping rate $\sigma_v$ (unit s$^{-1}$), as usual.
The  voids are considered as spherical-shaped extended defects (radius $r_0$) with a specific trapping rate $\alpha$ (unit m\,s$^{-1}$) which is related to the surface area of the void. In units of s$^{-1}$ the specific trapping rate of voids reads
\begin{equation}
\label{eq:sigma_t}
\sigma_t= \frac{\alpha 4\pi r_0^2}{\Omega},
\end{equation}
where $\Omega$ denotes the atomic volume.
 
The temporal and spatial evolution of the density  $\rho_l$ of free positrons in the lattice is governed by:
\begin{equation}
\label{differential_eq}
\frac{\partial \rho_l}{\partial t}=
D\nabla^{2}\rho_l-\rho_l\left(\frac{1}{\tau_f}+
\sigma_v C_v\right)
\end{equation}
where $C_v$ denotes the concentration of vacancy-type point defects in the matrix.
The positrons trapped in the voids are described in terms of their density $\rho_t$ obeying the rate equation
\begin{equation}
\label{eq:rho_t}
\frac{\mathrm{d}\rho_t}{\mathrm{d}t}=
\alpha \rho_l(r_{0},t)-\frac{1}{\tau_t}\rho_t.
\end{equation}
The temporal evolution of the number $N_v$ of $e^+$ trapped in the point defects in the lattice is given by
\begin{equation}
\frac{\mathrm{d}N_v}{\mathrm{d}t}=
-\frac{1}{\tau_v} N_v+\sigma_v C_v N_f \, ,
\end{equation}
where the number $N_f$ of positrons in the free state follows from integration of $\rho_l$:
\begin{equation}
N_f=\int \rho_l \mathrm{d}V.
\end{equation}

The continuity of the ${\rm e}^{+}$ flux at the boundary between the lattice and the void surface is expressed by\footnote{Note the negative sign in contrast to the model of $e^+$ trapping at grain boundaries (e.g., \cite{Oberdorfer09}) where the corresponding continuity equation refers to the outer boundary.}
\begin{equation}
D\nabla \rho_l\Big|_{r=r_{0}}-\alpha \rho_l(r_{0},t)=0.
\end{equation}

The outer radius $R$ of the diffusion sphere is related to the void concentration 
\begin{equation}
\label{eq:C_t}
C_t = \frac{3 \Omega}{4 \pi R^3} \, .
\end{equation}
The outer boundary condition
\begin{equation}
\frac{\partial \rho_l}{\partial r}\Big|_{r=R}=0
\end{equation}
reflects the vanishing $e^+$ flux through the outer border ($r = R$) of the diffusion sphere. 
This boundary condition is the same as applied earlier in a quite
different diffusion-reaction model of ortho-para conversion of positronium at reaction cerntres \cite{Wuerschum95}.

As initial condition we adopt the picture that at $t=0$ all thermalized positrons are in the free state and homogeneously distributed in the lattice, i.e., initial density $\rho_l = \rho_l (0)$,  $\rho_t(0)=0$, $N_v (0) = 0$.
Under this initial condition the solution of Eq. (\ref{differential_eq}) exhibits spherical symmetry.

Up to this point the above formulated diffusion-reaction problem is identical to that of Nieminen et al. \cite{Nieminen79}
apart from the additional rate-limited trapping at vacancy-type point defects which is considered here.
However, compared to \cite{Nieminen79}, in the following part of the present work the time dependence is handled by means of Laplace transformation which will lead to the more convenient
closed-form solutions. Applying the Laplace transformation 
\begin{eqnarray}
\nonumber
\tilde{\rho}_{l,t}(p)=\int\limits _{0}^{\infty}\exp(-pt)\rho_{l,t}(t)\mathrm{d}t,
\end{eqnarray}
\begin{equation}
\tilde{N}_{v,f}(p)=\intop_{0}^{\infty}\exp(-pt)N_{v,f}(t)\mathrm{d}t
\end{equation}
leads to the basic equations
\begin{equation}
\label{differential_eq_r}
\frac{\mathrm{d^{2}}\tilde{\rho}_l}{\mathrm{d}r^{2}}+\frac{2}{r}
\frac{\mathrm{d}\tilde{\rho}_l}{\mathrm{d}r}-\gamma^{2}\tilde{\rho}_l
=-\frac{\rho_l(0)}{D}
\end{equation}
with
\begin{eqnarray}
\label{eq:gamma}
\gamma^{2}=\gamma^{2}(p)=\frac{\tau_f^{-1}+\sigma_v C_v+p}{D} \, ,
\end{eqnarray}
and
\begin{equation}
\label{tilde_rho_t}
\tilde{\rho}_t=
\frac{\alpha \tilde{\rho}_l(r_{0},p)}{
\tau_{t}^{-1}+p} \: ,
\end{equation}
\begin{equation}
\label{N_v}
\tilde{N}_v=
\frac{\sigma_v C_v}{\tau_v^{-1}+p} \times \int\limits^{R}_{r_0} 4 \pi r^2 \tilde{\rho_l}(r,p) {\rm d} r\: ,
\end{equation}
with the boundary conditions
\begin{equation}
\label{boundary_condition1}
D\,\frac{\mathrm{d}\tilde{\rho}_l}{\mathrm{d}r}\Bigg|_{r=r_{0}}-
\alpha \tilde{\rho}_l(r_{0},p)=0
\end{equation}
and
\begin{equation}
\label{boundary_condition2}
\frac{\mathrm{d}\tilde{\rho}_l}{\mathrm{d}r}\Bigg|_{r=R}=0 \, .
\end{equation}

The solution of the differential equation (\ref{differential_eq_r}) satisfying equations
[Eq. (\ref{boundary_condition1})] and [Eq. (\ref{boundary_condition2})] can be written as
\begin{equation}
\label{solution_rho_tilde}
\tilde{\rho}_l(r,p)=A\, i_{0}^{(1)}(\gamma r)+ B\, i_{0}^{(2)}(\gamma r) +
\frac{\rho_l(0)}{\tau_f^{-1}+\sigma_v C_v + p}
\end{equation}
with
\begin{eqnarray}
\nonumber
A:=& \displaystyle{\alpha~\frac{\rho_l(0)}{\tau_f^{-1}+\sigma_v C_v+p}~\times~\frac{i_{1}^{(2)}(\gamma R)}{-D \gamma F_1 + \alpha F_2}} \, , \\
\label{AB}
B:= & \displaystyle{\alpha~\frac{\rho_l(0)}{\tau_f^{-1}+\sigma_v C_v+p}~\times~\frac{i_{1}^{(1)}(\gamma R)}{D \gamma F_1 - \alpha F_2}}
\end{eqnarray}
and
\begin{eqnarray}
\nonumber
F1= i_1^{(2)}(\gamma r_0) i_1^{(1)}(\gamma R) - i_1^{(1)}(\gamma r_0)  i_1^{(2)}(\gamma R) \, , \\
F2= i_0^{(2)}(\gamma r_0)  i_1^{(1)}(\gamma R) - i_0^{(1)}(\gamma r_0)  i_1^{(2)}(\gamma R) \, .
\end{eqnarray}

$i_{n}^{(1)}$ and $i_{n}^{(2)}$ ($n = 0,1$) denote the modified spherical Bessel functions of order $n$
\cite{olver2010nist}
\begin{eqnarray}
\nonumber
i_{n}^{(1)}(z):=(\frac{\pi}{2z})^{1/2}I_{n+1/2}(z), \,     
\end{eqnarray}
\begin{equation}
i_{0}^{(1)}=\frac{\sinh z}{z},\quad                        
i_{1}^{(1)}=\frac{\cosh z}{z}-\frac{\sinh z}{z^2}
\end{equation}
\begin{eqnarray}
\nonumber
i_{n}^{(2)}(z):=(\frac{\pi}{2z})^{1/2}I_{-n-1/2}(z), \,    
\end{eqnarray}
\begin{equation}
i_{0}^{(2)}=\frac{\cosh z}{z},\quad                        
i_{1}^{(2)}=\frac{\sinh z}{z}-\frac{\cosh z}{z^2}
\end{equation}
where $I_{\pm n\pm 1/2}(z)$ represents the Bessel function.

Basis for analyzing positron annihilation experiments is the total probability $n(t)$ that a
$e^+$ implanted at $t=0$ has not yet been annihilated at time $t$.
Here $n(t)$ is given by the number density of $e^+$ per lattice sphere at time
$t$:
\begin{equation}
n(t)= \frac{1}{\frac{4}{3}\pi (R^3 - r_{0}^{3})\rho_l(0)} \times
\left\{
\int\limits _{r_{0}}^{R}4\pi r^{2}\rho_l(r,t)\mathrm{d}r
+4\pi r_{0}^{2}\rho_t (t) + N_v (t)
\right\}       \:.
\end{equation}
The Laplace transform of $n(t)$ can be calculated taking into account the solution of  $\tilde{N}_v$ [Eq. (\ref{N_v})]
and the solution of the differential equation (\ref{solution_rho_tilde}) which yields
\begin{equation}
\tilde{n}(p)=\frac{1}{\frac{4}{3}\pi (R^3-r_{0}^{3})\rho_l(0)}
\times
\left\{
 \left(1+\frac{\sigma_v C_v}{\tau_v^{-1}+p}\right)
 \int\limits^{R}_{r_0} 4\pi r^{2} \tilde{\rho}_l(r,p) {\rm d} r
+ 4\pi r^{2}_{0} \tilde{\rho}_t(p)
\right\}       \:.
\end{equation}

Solving the integral after substituting $\tilde{\rho_t}(p)$ by Eq. (\ref{tilde_rho_t}),  
insertion of $A$ and $B$ [Eq. (\ref{AB})],
yields after some algebra

\begin{equation}
\label{Laplace_n}
\tilde{n}(p)=\frac{1}{t_{fc}^2 t_{v}  t_{t}} \Biggl\{t_{vc} t_{fc}  t_{t} + \frac{K (t_{fc}  t_{v} - t_{vc}  t_{t})
\Bigl(\gamma \hat{R} - \tanh(\gamma \hat{R}) [1- \gamma^2 r_0 R]\Bigr)}
{\gamma \hat{R}- \tanh(\gamma \hat{R}) [1-\gamma^2 r_0 R] + \frac{\alpha r_0}{D}[\gamma R - \tanh(\gamma \hat{R})]} \Biggr\} 
\end{equation}
with 
\begin{equation}
\label{eq:K}
K=\frac{3\alpha r_0^2}{R^3-r_0^3} \, ,
\end{equation}
\begin{equation}
\label{eq:R}
\hat{R} = R-r_0 \, ,
\end{equation}
and the abbreviations
\begin{eqnarray}
\nonumber
t_{t}=\tau_t^{-1}+p \, ; & t_{v}=\tau_v^{-1}+p \,; \\
t_{vc}=\tau_v^{-1}+\sigma C+p \, ; & t_{fc}=\tau_f^{-1}+\sigma C+p \, .
\end{eqnarray}
The Laplace transform $\tilde{n}(p)$ [Eq. (\ref{Laplace_n})] represents the
solution of the present diffusion and trapping model from which
both the mean positron lifetime and the positron lifetime spectrum
can be deduced.
The mean positron lifetime $\overline{\tau}$ is obtained  by
taking the Laplace  transform  at $p = 0$:
\begin{equation}
\label{tauq_n_tilde}
\overline{\tau} = \tilde{n}(p=0) =  \int\limits^{\infty}_{0}  n(t)
dt \, .
\end{equation}
The positron lifetime spectrum follows from  $\tilde{n}(p)$ by means of Laplace inversion. The single poles $p = - \lambda_i$
of  $\tilde{n}(p)$ in the complex $p$ plane define the decay rates $\lambda_i (i=0,1,2,\dots)$ of the positron lifetime spectrum:
\begin{equation}
\label{spectrum}
n(t) = \sum_{i=0}^{\infty}I_i \exp (-\lambda_i t)  \, ,
\end{equation}
where $I_i$ denote the relative intensities.

\section{Analysis}
At first, we consider the most important case that $e^+$ trapping  exclusively occurs at  voids, i.e., 
we omit $e^+$ trapping at point defects in the lattice ($C_v=0$).
For this case, we present the solution of the general diffusion-reaction theory (Sect.~\ref{sec:general}) and compare 
it with the limiting cases of entirely reaction-controlled trapping (Sect.~\ref{sec:rate_limit})
and entirely diffusion-controlled trapping (Sect.~\ref{sec:diffusion_limit}). Finally, the case of competitive 
reaction-controlled trapping at lattice defects is considered (Sect.~\ref{sec:vacancies})
and an extension to larger precipitates is presented for describing precipitate$-$matrix composite structures
(Sect.~\ref{sec:extended}).

\subsection{\label{sec:general}
General case with trapping at voids, exclusively ($C_v=0$)}

For negligible trapping at vacancies within the lattice ($C_v=0$), 
the diffusion-reaction model according to Eq.~(\ref{Laplace_n})  yields
for positron trapping in voids as the single type of trap:
\begin{equation}
\label{eq:n}
\tilde{n}(p) = \frac{1}{\tau_f^{-1}+p} \Biggl\{ 1+\frac{K(\tau_f^{-1}-\tau_t^{-1})}{(\tau_t^{-1}+p)(\tau_f^{-1}+p)} \times 
\frac{\gamma \hat{R}-\tanh(\gamma \hat{R}) [1-\gamma^2 r_0 R]}{\gamma \hat{R}-\tanh(\gamma \hat{R}) [1-\gamma^2 r_0 R]+ \frac{\alpha r_0}{D}[\gamma R- \tanh(\gamma \hat{R})]}
 \Biggr \} 
\end{equation}
and, hence, for the mean positron lifetime 
\begin{equation}
\label{eq:tauq}
\overline{\tau}=\tilde{n}(0)= \tau_f \Biggl\{1+ K (\tau_t-\tau_f) \times 
\frac{\gamma_0 \hat{R}-\tanh(\gamma_0 \hat{R}) [1-\gamma_0^2 r_0 R]}{\gamma_0 \hat{R}-\tanh(\gamma_0 \hat{R}) [1-\gamma_0^2 r_0 R]+ \frac{\alpha r_0}{D}[\gamma_0 R- \tanh(\gamma_0 \hat{R})]}
 \Biggr \} \, .
\end{equation}
The pole of Eq.~(\ref{eq:n}) for $p = - \tau_t^{-1}$ corresponds to the positron lifetime  component $\tau_t$ 
of the void-trapped state for which the following intensity is obtained:
\begin{equation}
\label{eq:I_t}
I_t= \frac{K}{\tau_f^{-1}-\tau_t^{-1}}\times 
\frac{\gamma_t \hat{R}-\tanh(\gamma_t \hat{R}) [1-\gamma_t^2 r_0 R]}{\gamma_t \hat{R}-\tanh(\gamma_t \hat{R}) [1-\gamma_t^2 r_0 R]+ \frac{\alpha r_0}{D}[\gamma_t R- \tanh(\gamma_t \hat{R})]} \,.
\end{equation}
In equations (\ref{eq:n}), (\ref{eq:tauq}), (\ref{eq:I_t}):
\begin{equation}
\label{eq:gamma_0_t}
\gamma^2=\frac{\tau _f^{-1}+p}{D}; \, 
 \gamma_0^2=\frac{\tau _f^{-1}}{D}; \, 
 \gamma_{t}^2=\frac{\tau _f^{-1}-\tau_{t}^{-1}}{D} \, .
\end{equation}

In addition to the annihilation component $\tau_t^{-1}$ of the void-trapped state, 
$\tilde{n}(p)$ [Eq.~(\ref{eq:n})] comprises a sequence of first-order poles $p = - \lambda_{0,j}$ for 
$\lambda_{0,j} > \tau_f^{-1}$. These  components $\lambda_{0,j}$,  
which define the fast decay rates ($\lambda_{0,j} > \tau_f^{-1}$) of the $e^+$ lifetime spectrum, 
are given by the solutions of the transcendental equation 
\begin{equation}
\tan(\gamma^{\star} \hat{R}) = \frac{\gamma^{\star} (\alpha r_0 R + D \hat{R})}{D(1+\gamma^{\star 2} r_0 R) + \alpha r_0}
\label{eq:transcendent}
\end{equation}
with
\begin{equation}
\gamma^{\star 2} = \frac{\lambda_{0,j}-\tau_f^{-1}}{D} 
\label{eq:gamma'}
\end{equation}
in agreement with the aforementioned earlier work of Nieminen et al. \cite{Nieminen79}.\footnote{Eq. (\ref{eq:transcendent}) 
is identical to the corresponding eq. (15) in the work of Nieminen et al. when $\nu$ in \cite{Nieminen79}
is identified with $4\pi r_0^2 \alpha$.}$^,$\footnote{We note that   
the same problem was treated in the framework of a more general theoretical approach by K\"ogel \cite{Koegel96}. The 
quoted specific function  in dependence of $\gamma \hat{R}$ [eq. (75) in \cite{Koegel96}], which
determines the mean $e^+$ lifetime and the intensity of the trap component,
 however, is not readily applicable.
}
As usual for this kind of diffusion-reaction problem (see, e.g. \cite{Oberdorfer09}),  
the intensities of these decay rates rapidly decrease.
Experimentally only a single fast decay rate can be resolved in addition to the decay rate $\tau_t^{-1}$ of the trapped state. An experimental two-component $e^+$ lifetime spectrum is practically entirely  defined by 
$\overline{\tau}$ [Eq.~(\ref{eq:tauq})] and by $\tau_t$ with the corresponding intensity $I_t$ [Eq.~(\ref{eq:I_t})].

The appearance of a second-order pole in Eq.~(\ref{eq:n}) at $p = - \tau_f^{-1}$ (i.e., $\gamma = 0$) is spurious. Closer inspection by applying Taylor expansion shows that the intensity associated with this pole cancels.

Following the consideration of Dryzek \cite{dryzek2002}, in analogy to the mean $e^+$ lifetime [Eq. (\ref{eq:tauq})]
a respective relation for the mean line shape parameter $\overline{S}$ of 
Doppler broadening of the positron-electron annihilation
can be given:
\begin{equation}
\label{eq:Doppler}
\overline{S}=  S_f \Biggl\{1+ K (S_t-S_f) \times 
\frac{\gamma_0 \hat{R}-\tanh(\gamma_0 \hat{R}) [1-\gamma_0^2 r_0 R]}{\gamma_0 \hat{R}-\tanh(\gamma_0 \hat{R}) [1-\gamma_0^2 r_0 R]+ \frac{\alpha r_0}{D}[\gamma_0 R- \tanh(\gamma_0 \hat{R})]}
 \Biggr \} \, ,
\end{equation}
where $S_f$ and $S_t$ denote the line shape parameters of the free and trapped state, respectively.

For the sake of completeness, we quote $\tilde{n}(p)$ without derivation for the case that 
at time zero positrons are homogeneously distributed in the voids and the lattice, 
i.e., for the initial condition $\rho_t (0) = r_0 \rho_l (0)/3$: 
\begin{eqnarray}
\label{eq:n_homogen}
\nonumber
\tilde{n}(p) =  \displaystyle{\frac{1}{\tau_f^{-1}+p} \Biggl\{ 1+ \frac{r_0^3}{R^3} \times
\frac{\tau_f^{-1}-\tau_t^{-1}}{\tau_t^{-1} + p} +}
 \frac{3 \alpha r_0^2}{R^3} \times \frac{\tau_f^{-1}-\tau_t^{-1}}{(\tau_t^{-1}+p)(\tau_f^{-1}+p)} 
\times \\
 \displaystyle{\frac{\gamma \hat{R}-\tanh(\gamma \hat{R}) [1-\gamma^2 r_0 R]}{\gamma \hat{R}-\tanh(\gamma \hat{R}) [1-\gamma^2 r_0 R]+ \frac{\alpha r_0}{D}[\gamma R- \tanh(\gamma \hat{R})]}
 \Biggr \}} \, .
\end{eqnarray}
Eq.~(\ref{eq:n_homogen}) includes in the limiting case  of negligible trapping ($\alpha = 0$)
as  mean $e^+$ lifetime $\overline{\tau} = \tilde{n}(0) = 
[(R^3-r_0^3) \tau_f + r_0^3 \tau_t]/R^3$ the expected volume-averaged mean value of $\tau_f$ and $\tau_t$.

\subsection{\label{sec:rate_limit}
Limiting case of entirely reaction limited trapping ($C_v=0$)}

If the e$^+$ diffusivity is high ($\gamma \hat{R} \ll 1$), the hyperbolic tangent 
in Eq.~(\ref{eq:n})  can be expanded.
Expansion up to the third order 
\begin{equation}
\tanh(z)\approx z-\frac{z^3}{3}
\end{equation}   
yields the mean $e^+$ lifetime 
\begin{equation}
\label{eq:tauq_rate}
\overline{\tau}=\displaystyle{
\tau_f \frac{1+ K \tau_t}{1+K \tau_f}}
\end{equation}
and for the $e^+$ lifetime component $\tau_t$ the intensity 
\begin{equation}
\label{eq:I_t_rate}
I_t=\displaystyle{
\frac{K }{\tau_f^{-1}+ K -\tau_t^{-1}}
}
\end{equation}
with $K$ according to Eq.~(\ref{eq:K}).
Equations (\ref{eq:tauq_rate}) and (\ref{eq:I_t_rate}) are the well-known solutions of the simple trapping model
when we identify $K$ for vanishing defect volume with the trapping rate $\sigma_t C_t$ 
[equations (\ref{eq:sigma_t}) and (\ref{eq:C_t})].
Note that the standard trapping model does not take into account the finite defect volume (here $4 \pi r_0^3/3$)
and, therefore, does not contain the subtrahend $r_0^3$ as in Eq. (\ref{eq:K}). With this subtrahend,
equations (\ref{eq:tauq_rate}) and (\ref{eq:I_t_rate}) correctly contain 
the exact values  $\overline{\tau} = \tau_t$ and $I_t = 1$ 
as limiting case for $R=r_0$.

\subsection{\label{sec:diffusion_limit}
Limiting case of entirely diffusion limited trapping ($C_v=0$)}

The present solution includes in the limiting special case $\alpha \rightarrow \infty$ the relationships
for an entirely diffusion-limited trapping, i.e., for Smoluchowski-type boundary condition 
\begin{equation}
\label{eq:Smoluchowski}
\rho_l (r_0, t) = 0 \, .
\end{equation}
In this limit one obtains from the Laplace transform [Eq. (\ref{eq:n})]
the mean $e^+$ lifetime 
\begin{equation}
\label{eq:tauq_diffusion}
\overline{\tau}=\tau_f \Biggl \{ 1+ \frac{3r_{\rm 0}D}{R^3-r_{\rm 0}^3}(\tau_t-\tau_f) 
\frac{\gamma_0 \hat{R}-\tanh(\gamma_0 \hat{R})[1-\gamma_0^2 r_0 R]}{\gamma_0 R-\tanh(\gamma_0 \hat{R})} \Biggr\}
\end{equation}
and for the trap component $\tau_t$ the intensity
\begin{equation}
\label{eq:I_t_diffusion}
I_t=\frac{3 r_0 D}{R^3-r_0^3} \times \frac{1}{\tau_f^{-1}-\tau_t^{-1}} \times \frac{\gamma_t \hat{R}-\tanh(\gamma_t \hat{R})[1-\gamma_t^{2}r_0 R]}{\gamma_t R-\tanh(\gamma_t \hat{R})}
\end{equation}
with $\gamma_0$, $\gamma_t$ according to eq. (\ref{eq:gamma_0_t}).

 \subsection{\label{sec:vacancies}
General case with voids {\em and} lattice vacancies}

The positron annihilation characteristics of diffusion-reaction
controlled trapping at voids and concomitant transition-limited
trapping at point defects in the lattice is given by Eq.
(\ref{Laplace_n}) in combination with Eq. (\ref{tauq_n_tilde}) and
Eq. (\ref{spectrum}).
The mean positron lifetime [Eq. \ref{tauq_n_tilde}], obtained from
Eq. (\ref{Laplace_n}) for $p=0$,  reads in the general case:
\begin{eqnarray}
\label{eq:tau_q_general}
\nonumber
\overline{\tau}=  \displaystyle{
\frac{1}{(\tau_f^{-1}+\sigma_v C_v)^{2}} }
\Biggl\{(\tau_f^{-1}+\sigma_v C_v)  (\tau_v^{-1}+\sigma_v C_v) \tau_v+  \\
\displaystyle{
\frac{K \Bigl((\tau_f^{-1}+\sigma_v C_v)  \tau_{t} - (\tau_v^{-1}+\sigma_v C_v)  \tau_{v} \Bigr)\Bigl(\gamma_0 \hat{R} - \tanh(\gamma_0 \hat{R})[1- \gamma^2 r_0 R]\Bigr)}
{\gamma_0 \hat{R}- \tanh(\gamma_0 \hat{R})[1-\gamma_0^2 r_0 R] + \frac{\alpha r_0}{D}[\gamma_0 R - \tanh(\gamma_0 \hat{R})]}\Biggr\}
}
\end{eqnarray}
with
\begin{equation}
\label{eq:gamma_v}
\gamma_0^2=\frac{\tau_f^{-1}+\sigma_v C_v}{D} \, .
\end{equation}

In addition to the pole $p = - \tau_t^{-1}$ which characterizes the void trapped state,
$\tilde{n}(p)$ [Eq. (\ref{Laplace_n})] contains the further defect-related pole  
$p = - \tau_v^{-1}$ for the vacancy-type defect in the lattice.
From the residues of  $\tilde{n}(p)$ [Eq. (\ref{Laplace_n})], 
the corresponding relative intensities
\begin{equation}
\label{eq:I_t_general}
I_t=\frac{K}{\tau_f^{-1}+\sigma_v C_v-\tau_t^{-1}}\times\frac{\gamma_t \hat{R} - \tanh(\gamma_t \hat{R})[1- \gamma_t^2 r_0 R]}{\gamma_t \hat{R}- \tanh(\gamma_t \hat{R})[1-\gamma_t^2 r_0 R] + 
\frac{\alpha r_0}{D}[\gamma_t R - \tanh(\gamma_t \hat{R})]}
\end{equation}
and 
\begin{eqnarray} 
\nonumber
I_v=
\frac{\sigma_v C_v}{\tau_f^{-1}+\sigma_v C_v-\tau_v^{-1}}  \Biggl\{1- \frac{K}{\tau_f^{-1}+\sigma_v C_v-\tau_v^{-1}} \times \\
\label{eq:I_v} 
\frac{\gamma_v \hat{R} - \tanh(\gamma_v \hat{R})[1- \gamma_v^2 r_0 R]}{\gamma_v \hat{R}- \tanh(\gamma_v \hat{R})[1-\gamma_v^2 r_0 R] + \frac{\alpha r_0}{D}(\gamma_v R - \tanh(\gamma_v \hat{R})]}\Biggr\}
\end{eqnarray} 
are deduced with 
\begin{equation}  
\label{eq:gamma_t_v}
\gamma_{t,v}^2=\frac{\tau _f^{-1}+\sigma_v C_v-\tau_{t,v}^{-1}}{D} \, .
\end{equation}


\subsection{\label{sec:extended}
Extended model for larger preciptates  with  $e^+$-trapping from both sides of precipitate$-$matrix interface}
The model presented above describes $e^+$ annihilation from a trapped state ($\tau_t$)
in spherical defects. Particularly, for larger precipitate sizes a situation may prevail where 
$e^+$ annihilation inside the precipitates occurs from a free state with a characteristic $e^+$ lifetime $\tau_p$
and where also from this free precipitate state positrons may  get trapped into the spherical interfacial shell between the precipitate and the surrounding matrix. This means that the precipitates are characterized by two compoments, one corresponding to the precipitate volume ($\tau_p$) and one corresponding to the trapped state in the matrix$-$precipitate interface ($\tau_t$).

The present model can be extended in a straight forward manner to this case under the reasonable assumption that the $e^+$ trapping from inside the precipitates is entirely reaction controlled. This is pretty well fulfilled as long as
the precipitate diameter is remarkably lower than the $e^+$ diffusion length in the precipitate.\footnote{A further model extension avoiding this
constraint will be outlined below.} 
In this case the extension can be described by an additional rate equation for the temporal evolution of the number $N_p$ of $e^+$ 
inside the precipitates 
\begin{equation}
\label{eq:N_p}
\frac{\mathrm{d}N_p}{\mathrm{d}t}=
-\Bigl( \frac{1}{\tau_p} + \frac{3\beta}{r_0} \Bigr) N_p \, ,
\end{equation}
where $\beta$ denotes the specific trapping rate (in units of m/s) at the spherical interfacial shell.
This trapping from inside the precipitates, which occurs in addition to the diffusion- and reaction-limited trapping
into the interfacial shell from the surrounding matrix, has to be taken into account in the rate equation for 
$\rho_t$ (Equation \ref{eq:rho_t})
by the additional summand $\beta \rho_p(t)$ with the number density
$\rho_p = 3 N_p / (4 \pi r_0^3)$ of $e^+$ in the precipitate.

Assuming a homogeneous distribution of $e^+$ at time zero  in the matrix and the 
precipitate ($\rho_l (0) = \rho_p (0)$) without  $e^+$ in the trapped state ($\rho_t (0) = 0$) for $t=0$,
one obtains with the Laplace transform of eq.~(\ref{eq:N_p})
\begin{equation}
\label{eq:Laplace_N_p}
\tilde{N_p} = \displaystyle{\frac{N_p(0)}{\tau_p^{-1} + \frac{3 \beta}{r_0} + p}}
\end{equation}
the additional summand
\begin{equation}
\label{eq:n_extension}
\Bigl( \frac{r_0}{R} \Bigr)^3 
\displaystyle{\Bigl(\frac{\frac{3 \beta}{r_0}}{\tau_t^{-1} +p} +1 \Bigr)
\frac{1}{\tau_p^{-1} + \frac{3 \beta}{r_0} + p}}
\end{equation}
in  eq. (\ref{Laplace_n}) of $\tilde{n}(p)$. 
Moreover, in the bracket of eq. (\ref{Laplace_n}) the first summand  is extended by the weighting factor 
$[ 1 - (r_0/R)^3]$ and  the  trapping rate $K$ (Equation \ref{eq:K}) in the second summand is replaced by 
$3 \alpha r_0^2/R^3$.

For \underline{$C_v = 0$} this leads to the mean $e^+$ lifetime 
\begin{eqnarray}
\label{eq:tauq_extended}
&\overline{\tau}= \tau_f \Biggl\{\displaystyle{ \Bigl[1 - \Bigl(\frac{r_0}{R} \Bigr)^3 \Bigr]} + 
\nonumber \\
& \displaystyle{\frac{3 \alpha r_0^2}{R^3} 
\times (\tau_t-\tau_f) \times 
\frac{\gamma_0 \hat{R}-\tanh(\gamma_0 \hat{R}) [1-\gamma_0^2 r_0 R]}{\gamma_0 \hat{R}-\tanh(\gamma_0 \hat{R}) [1-\gamma_0^2 r_0 R]+ \frac{\alpha r_0}{D}[\gamma_0 R- \tanh(\gamma_0 \hat{R})]} \Biggr \} } + \nonumber \\
& \displaystyle{\Bigl(\frac{r_0}{R} \Bigr)^3 \times \tau_t \times \frac{\tau_t^{-1} + \frac{3 \beta}{r_0}}{\tau_p^{-1} + \frac{3 \beta}{r_0}}}
\, , 
\end{eqnarray}
as compared to eq. (\ref{eq:tauq}).
Eq.~(\ref{eq:tauq_extended}) includes in the limiting case  of negligible trapping ($\alpha = \beta = 0$)
as  mean $e^+$ lifetime $\overline{\tau} = 
[(R^3-r_0^3) \tau_f + r_0^3 \tau_p]/R^3$ the expected volume-averaged mean value of $\tau_f$ and $\tau_p$.  

The additional pole for $p = - (\tau_p^{-1} + 3 \beta /r_0)$ of  $\tilde{n}(p)$ yields the intensity 
of the $e^+$ lifetime component $\tau_p$ in the precipitate:
\begin{equation}
I_p = \Bigl(\frac{r_0}{R} \Bigr)^3 \times 
\Biggl(1 - \displaystyle{\frac{\frac{3 \beta}{r_0}}{\tau_p^{-1} + \frac{3 \beta}{r_0} - \tau_t^{-1}} \Biggr)}
\, . 
\label{eq:I_p}
\end{equation}
Apart from the weighting prefactor ($(r_0/R)^3$), $I_p$ corresponds to the solution of the simple trapping model.\footnote{
Note that $I_p$ characterizes the free state in the precipitate.} Without trapping ($\beta = 0$), $I_p$ simply takes  the form of the weighting prefactor $(r_0/R)^3$.

Since $e^+$ trapping into the precipitate$-$matrix interface occurs both from inside the precipitate and from the surrounding matrix, the intensity  
of the trap component $\tau_t$ is given by the sum
\begin{equation}
I_t = I_t^{precip} + I_t^{matrix} \text{ with } 
I_t^{precip} = \Bigl(\frac{r_0}{R} \Bigr)^3 \times 
\displaystyle{\frac{\frac{3 \beta}{r_0}}{\tau_p^{-1} + \frac{3 \beta}{r_0} - \tau_t^{-1}} }
\, ,
\label{eq:I_t_tot}
\end{equation}
where $I_t^{matrix}$ corresponds to the intensity $I_t$ according to eq.~(\ref{eq:I_t}) with  
$K$   replaced by $3 \alpha r_0^2/R^3$.\footnote{The identical equation for $I_t$ (equation~\ref{eq:I_t_tot})  follows 
from the root $p = - \tau_t$ of the Laplace transform $\tilde{n}(p)$ in which the above mentioned extensions 
of eq. (\ref{Laplace_n}) are taken into consideration.}
 
We note that the two $e^+$ trapping processes into the precipitate$-$matrix interface, namely that from inside the precipitate and that from the surrounding matrix, are completely decoupled. The trapping process from inside the precipitate can, therefore, be treated independently.
This also means that the process has not to be restricted to the case of entirely reaction-controlled trapping as given above, but 
that $e^+$ trapping at the precipitate$-$matrix interface from inside the spherical precipitates can also be treated 
in the framework of diffusion-reaction theory. Hence, the available solutions for 
diffusion- and reaction-limited trapping at grain boundaries (GBs) of spherical crystallites \cite{Wuerschum96, Oberdorfer09} can 
be directly applied. For this purpose the solutions for the GB-model have simply to be weighted by the factor $(r_0/R)^3$ which denotes the volume fraction of the precipitates.\footnote{Given the above initial condition $\rho_l (0) = \rho_p (0)$ and $\rho_t (0) = 0$.}

For instance, for the mean $e^+$ lifetime, the last summand in Eq.~(\ref{eq:tauq_extended}), i.e., the rate-equation solution has 
to be replaced by that calculated for diffusion- and reaction-limited trapping at GBs \cite{Wuerschum96, Oberdorfer09},
yielding:
\begin{eqnarray}
\label{eq:tauq_double_extended}
&\overline{\tau}= \tau_f \Biggl\{\displaystyle{ \Bigl[1 - \Bigl(\frac{r_0}{R} \Bigr)^3 \Bigr]} + 
\nonumber \\
& \displaystyle{\frac{3 \alpha r_0^2}{R^3} 
\times (\tau_t-\tau_f) \times 
\frac{\gamma_0 \hat{R}-\tanh(\gamma_0 \hat{R}) [1-\gamma_0^2 r_0 R]}{\gamma_0 \hat{R}-\tanh(\gamma_0 \hat{R}) [1-\gamma_0^2 r_0 R]+ \frac{\alpha r_0}{D}[\gamma_0 R- \tanh(\gamma_0 \hat{R})]} \Biggr \} } + \nonumber \\
& \displaystyle{\Bigl(\frac{r_0}{R} \Bigr)^3 \Biggl\{ 
\tau_p +  (\tau_t-\tau_p) \times 
\frac{3 \beta L(\gamma'_0 r_0)}{r_0 \gamma'_0
\Bigl( \beta + \gamma'_0 D L (\gamma'_0 r_0) \Bigr) }\Biggr\}}
\, , 
\end{eqnarray}
with $\gamma'_0 = (\tau_p D)^{-1/2}$,  $\gamma_0 = (\tau_f D)^{-1/2}$
 and the Langevin function
\begin{equation}
L (z) = \coth z - \frac{1}{z} \, .
\label{eq:langevin}
\end{equation}

Likewise the intensity component $I_t^{precip}$
 of the rate-equation solution in Eq.~(\ref{eq:I_t_tot}) has to be replaced by \cite{Wuerschum96, Oberdorfer09}:
\begin{equation}
\label{eq:I_p_diff}
I_t^{precip} = \Bigl(\frac{r_0}{R} \Bigr)^3 \times \frac{3 \beta}{r_0(\tau_p^{-1}-\tau_t^{-1})} \times
\Biggl\{
\frac{\gamma'_t D L(\gamma'_t r_0)}{
 \beta + \gamma'_t D L(\gamma'_t r_0)  }
\Biggr\}
\, 
\end{equation}
with 
\begin{equation}
\gamma'^2_t = \frac{\tau_p^{-1} - \tau_t^{-1}}{D} \, .
\label{eq:gamma'_t}
\end{equation}
For the sake of completeness we quote the mean $e^+$ lifetime for reaction-controlled trapping from both in- and outside:
\begin{equation}
\label{eq:tauq_extended_rate}
\overline{\tau}= \displaystyle{\tau_t \Bigl(1 - \Bigl[\frac{r_0}{R} \Bigr]^3 \Bigr) 
\frac{\tau_t^{-1} + \frac{3 \alpha r_0^2}{R^3-r_0^3}}{\tau_f^{-1} + \frac{3 \alpha r_0^2}{R^3-r_0^3}}
+ \tau_t \Bigl(\frac{r_0}{R} \Bigr)^3  \, \, \frac{\tau_t^{-1} + \frac{3 \beta}{r_0}}{\tau_p^{-1} + \frac{3 \beta}{r_0}}}
\, . 
\end{equation}
A further extension for taken into account additional $e^+$ trapping at point defects
inside the matrix (Sect.~\ref{sec:vacancies})
and inside the precipitates (in analogy to the GB model \cite{Oberdorfer09}) is straightforward, so that 
the corresponding equations have not to be stated explicitly.

\section{\label{discussion} Discussion}
\subsection{\label{sec:voids}
Voids, clusters, small precipitates}

The presented model with the exact solution of diffusion-reaction
controlled trapping at voids (or other extended spherical defects like clusters and small precipitates)
and competitive transition-limited
trapping at vacancy-type defects yields closed-form expressions
for the mean positron lifetime $\overline{\tau}$ [Eq. (\ref{eq:tau_q_general})]
and for the relative intensities $I_t$ [Eq. (\ref{eq:I_t_general})]
and $I_v$ [Eq. (\ref{eq:I_v})] of the $e^+$ lifetime components
$\tau_t$ and $\tau_v$ of the void and the vacancy
trapped states, respectively.

\begin{figure}[ht]
\includegraphics[width=8.5cm]{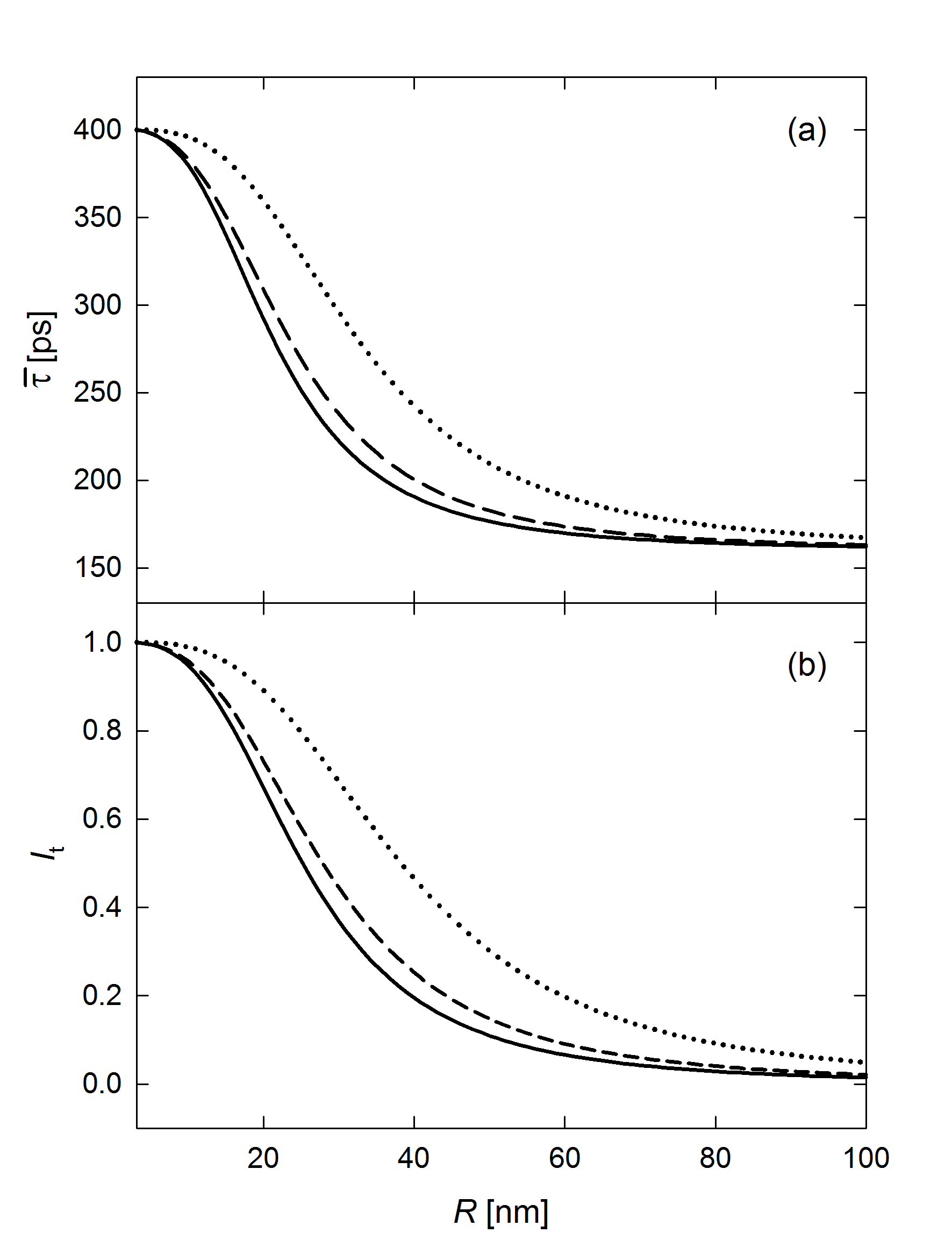}
\caption{(a) Mean $e^+$ lifetime $\overline{\tau}$
and (b) relative intensity $I_t$ of void component $\tau_t$
in dependence of diffusion radius $R$
for diffusion-reaction model (---) [Equations (\ref{eq:tauq}), (\ref{eq:I_t})], 
for standard rate ($---$) [Equations (\ref{eq:tauq_rate}), (\ref{eq:I_t_rate})], and for limiting
case of entirely diffusion-limited trapping ($\cdot \cdot \cdot$) [Equations (\ref{eq:tauq_diffusion}), (\ref{eq:I_t_diffusion})].
Parameters: $\tau_f=160$~ps, $\tau_t=400$~ps,
$D=2 \times 10^{-5}$~m$^2$s$^{-1}$, $\alpha=3 \times 10^3$~ms$^{-1}$,
$r_0 = 3$~nm. Note that $R$ is related to the void concentration [Eq. \ref{eq:C_t}].
}
\label{fig:2}
\end{figure}

We start the discussion considering exclusively diffusion-reaction
controlled trapping at voids  (Sect. \ref{sec:general}).
The model contains as limiting cases both the solution of the simple trapping model 
(Sect. \ref{sec:rate_limit}) and the one of the entirely diffusion-limited trapping (Sect. \ref{sec:diffusion_limit}).
The mean $e^+$ lifetime $\overline{\tau}$ [Eq. (\ref{eq:tauq})] and the intensity 
$I_t$ [Eq. (\ref{eq:I_t})] in dependence of the radius $R$ of the diffusion sphere
are compared in Fig. \ref{fig:2} with the two limiting cases. Note, that $R$ is related to the the void concentration 
[Eq. (\ref{eq:C_t})].
For illustration the following characteristic e$^+$ annihilation parameters are used:
a free $e^+$ lifetime  $\tau_f=160$~ps as typical for aluminium, 
a $e^+$ lifetime $\tau_t=400$~ps as typical for voids \cite{Eldrup03},
a $e^+$ diffusion coefficient $D=2 \times 10^{-5}$~m$^2$s$^{-1}$, a void radius $r = 3$~nm, and  
a specific e$^+$ trapping rate  $\alpha = 3 \times 10^3$~ms$^{-1}$
reported by Dupasquier et al. \cite{Dupasquier93} for interfaces in Al. For surfaces of Al 
a value $\alpha = 7.6 \times 10^3$~ms$^{-1}$ was calculated by Nieminen and Lakkonnen \cite{Nieminen79a}.
Using an atomic volume $\Omega$ for Al of $\Omega^{-1} = 6 \times 10^{28}$~m$^{-3}$,   
$\alpha = 3 \times 10^3$~ms$^{-1}$ corresponds to  a trapping rate $\sigma_t = 2\times 10^{16}$ s$^{-1}$ 
[Eq. \ref{eq:sigma_t}] which is similar to that deduced by Bentzon and Evans \cite{Bentzon90} for voids in Mo.\footnote{
A value $\sigma_t = 4\times 10^{16}$ s$^{-1}$ is deduced from 
the trapping rate of $3.2 \times 10^9$~s$^{-1}$ at 300~K and 
a void number density of $5.3 \times 10^{21}$~m$^{-3}$
quoted in \cite{Bentzon90}.}

Both $\overline{\tau}$ (Fig. \ref{fig:2}a) and $I_t$ (Fig. \ref{fig:2}b) exhibit the characteristic sigmoidal increase
from the free state to the saturation-trapped state with decreasing $R$, i.e., increasing void concentration $C_t$.
Compared to the exact solution of the present model, the standard trapping model 
and the limiting case of entirely diffusion-limited trapping 
show qualitatively the same trend for $\overline{\tau}$ and $I_t$.
However, both special cases systematically overestimate $\overline{\tau}$ and $I_t$, i.e., predict 
stronger trapping since either the rate-limiting effect or the diffusion-limiting effect are neglected in these
approximations. For instance, if one would determine  the void concentration from a typical, experimentally measured intensity $I_t$ of
45\,\% \cite{Nambissan1989}, a concentration 36\,\% too low would be deduced from the standard trapping model 
compared to the exact theory for the parameter set according to Fig. \ref{fig:2}b.

The deviations of the two limiting cases from the exact solution become even more clear when 
the ratios of the trap component intensities of the limiting and exact solution is considered
as shown in the upper part of Fig. \ref{fig:3}.
The deviation from the exact solution substantially increases with decreasing intensity, i.e., with
decreasing void concentration. In this low concentration regime, the deviations 
attain a factor of ca. 1.5 (reaction limit) or larger than 3 (diffusion limit)
for the present set of parameters, i.e., the entirely diffusion-limiting case
deviates in this example more strongly than the reaction-limited case. 
Diffusion limitation gets even more pronounced when $e^+$ diffusivity is reduced, e.g., due to
scattering at lattice imperfections.
Regarding the opposite side of high defect concentrations,  
Fig. \ref{fig:3} (upper part) nicely demonstrates that deviations from the exact theory  
vanishes upon approaching $e^+$ saturation trapping since in this regime  
kinetic effects tends to become irrelevant.

\subsubsection{\label{sec:effective_rate} Comparison with effective rate approach}
Next we compare the present model with approximations according to which
diffusion limitation is taking account in the standard trapping model by means of a diffusion-limited trapping rate 
\cite{Seeger74,Bentzon90}:
\begin{equation}
K_{diff} = \frac{4\pi r_0 D}{\Omega} \times C_t \,.
\label{eq:K_diffusion}
\end{equation}
The case of both transition- and diffusion-limited trapping, is treated in this approximation by means of
the effective trapping rate \cite{Seeger74,Bentzon90}
\begin{equation}
K_{eff} = \frac{K_{diff} \sigma_t C_t}{K_{diff} + \sigma_t C_t} 
\label{eq:K_eff}
\end{equation}
with $\sigma_t$ and $C_t$ according to equations (\ref{eq:sigma_t}) and  (\ref{eq:C_t}), respectively.
We note that the diffusion-limited trapping rate according to eq. (\ref{eq:K_diffusion}) is also included in the present model; in fact $K_{diff}$ is identical to the pre-factor of $I_t$ for entirely diffusion limited trapping [Eq. (\ref{eq:I_t_diffusion})] when the 
subtrahend $r_0^3$ in the nominator, which is associated with the defect volume,  is omitted.

In figure \ref{fig:4} the concentration dependence of the relative intensity 
$I_t$ of the $e^+$ lifetime component $\tau_t$ 
in voids is shown for the exact models of diffusion-reaction  [Eq. (\ref{eq:I_t})] or 
entire diffusion limitation [Eq. (\ref{eq:I_t_diffusion})] in comparison with the corresponding approximations
using the above mentioned effective or  diffusion-trapping rates [Equations (\ref{eq:K_diffusion}), (\ref{eq:K_eff})] 
with the simple trapping  model [Eq. (\ref{eq:I_t_rate})].
Although the effective-rate approximations of the diffusion limitation describe the sigmoidal curve fairly well,
deviations from the exact diffusion models are also apparent, e.g., for the example, $I_t = 45$\,\%, mentioned above 
the deviation in concentration is ca. 7\,\% compared to the exact diffusion-reaction theory.

\begin{figure}[ht]
\includegraphics[width=8.5cm]{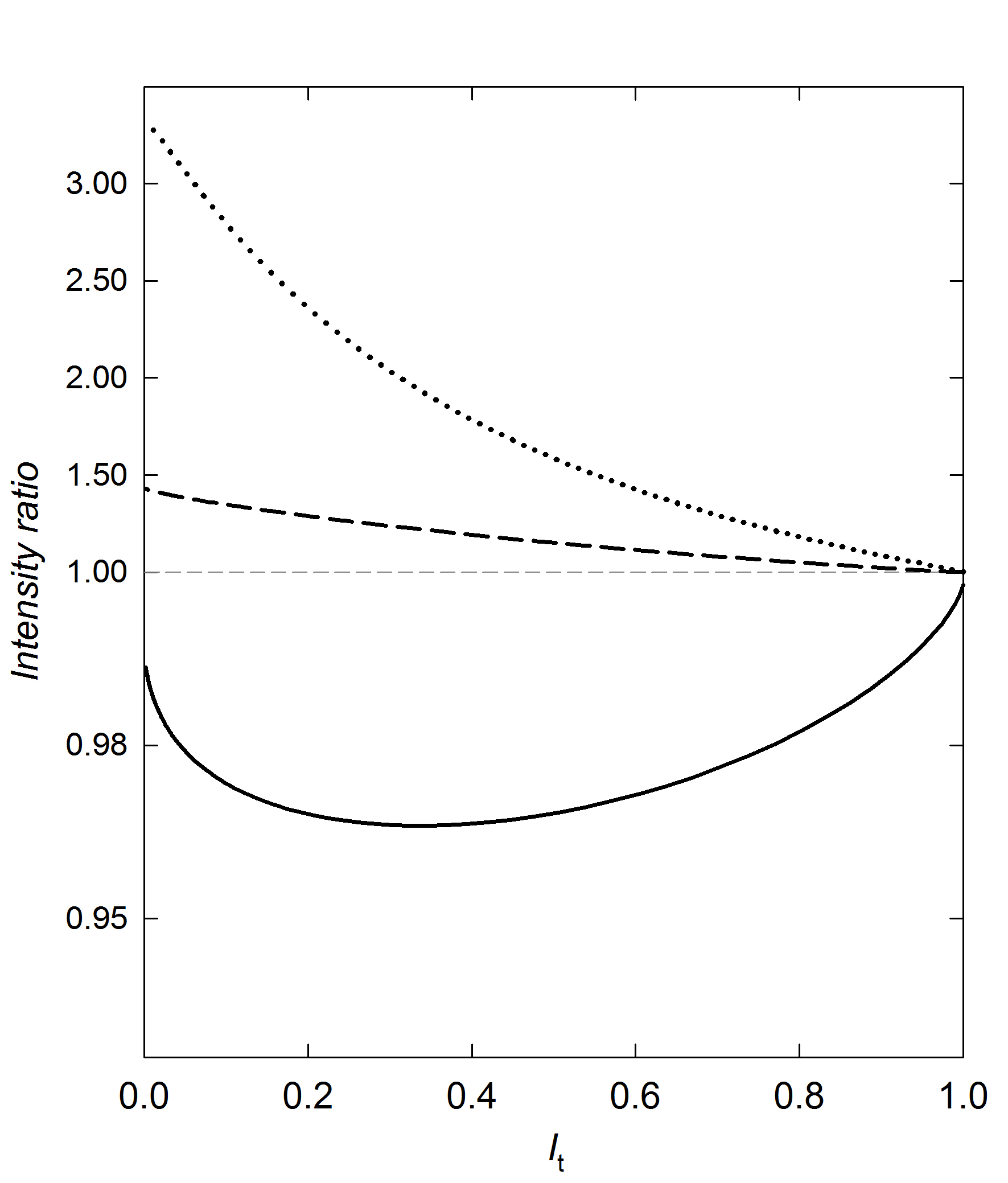}
\caption{Ratio of approximate intensity and exact intensity of void component $\tau_t$
in dependence of the exact intensity $I_t$ according to eq. (\ref{eq:I_t}).
Approximate intensity: entirely diffusion-limited trapping ($\cdot \cdot \cdot$) [Eq. (\ref{eq:I_t_diffusion})],
entirely reaction-controlled trapping  ($---$) [Eq. (\ref{eq:I_t_rate})], 
and simple trapping model [Eq. (\ref{eq:I_t_rate})] with effective diffusion-trapping rate 
$K_{eff}$ [Eq. (\ref{eq:K_eff})] (---). Note the different scales of the ratio axis
for ratios $< 1$ and $> 1$.
Parameters: 
$\Omega^{-1} = 6\times 10^{28}$~m$^{-3}$, others as in
Fig. \ref{fig:2}.
}
\label{fig:3}
\end{figure}

\begin{figure}[ht]
\includegraphics[width=8.5cm]{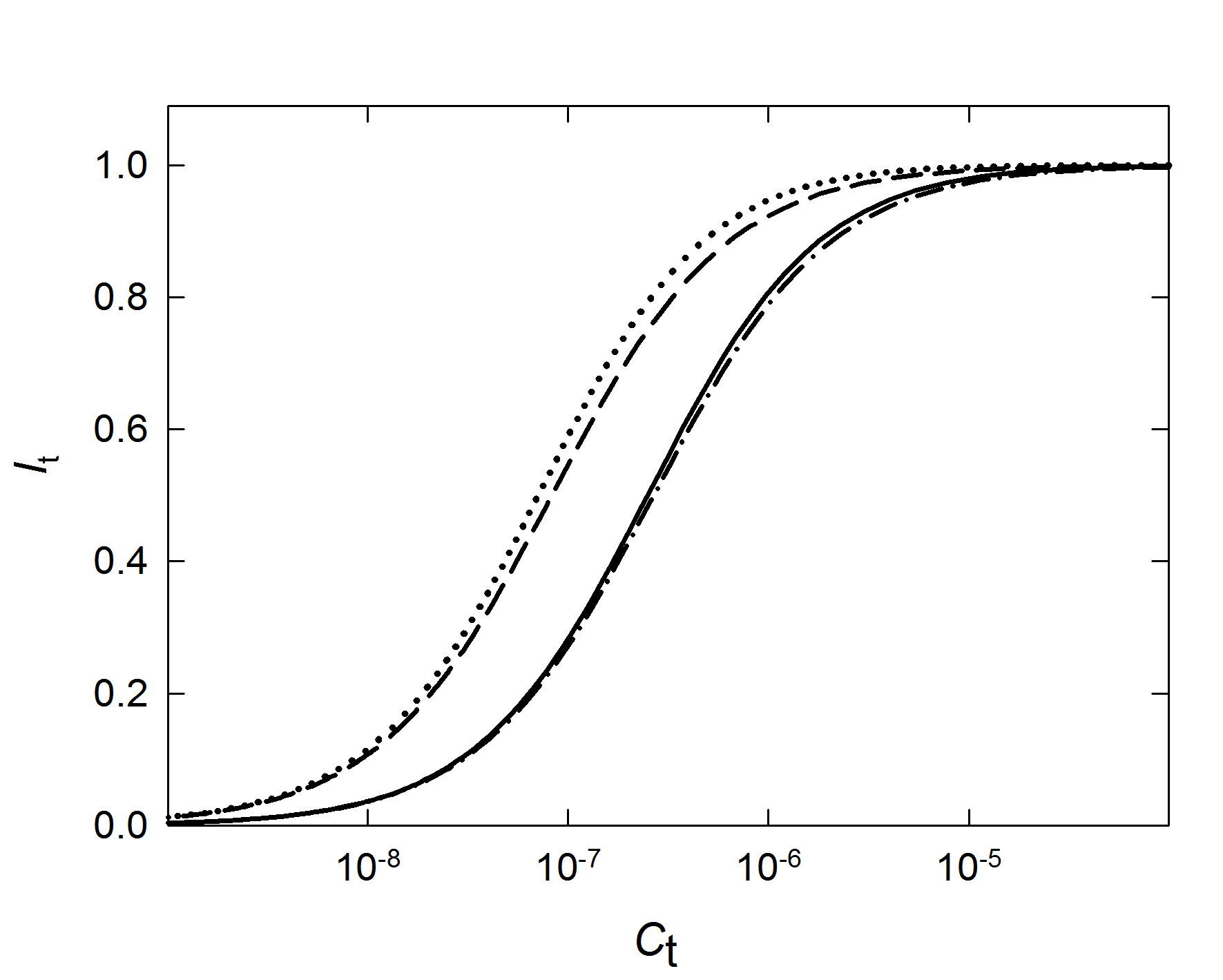}
\caption{
Relative intensity $I_t$ of void component $\tau_t$ 
in dependence of voids concentration $C_t$  
for diffusion-reaction model (---) [Eq. (\ref{eq:I_t})], 
for limiting case of entirely diffusion-limited trapping ($\cdot \cdot \cdot$) [Eq. (\ref{eq:I_t_diffusion})],
as well as for simple trapping model [Eq. (\ref{eq:I_t_rate})] with effective trapping rate of diffusion  
$K_{diff}$ [Eq. (\ref{eq:K_diffusion})] ($---$) or with effective diffusion- and transition-limited trapping rate 
$K_{eff}$ [Eq. (\ref{eq:K_eff})] ($- \cdot - \cdot -$).
Parameters: 
$\Omega^{-1} = 6\times 10^{28}$~m$^{-3}$, others as in
Fig. \ref{fig:2}.
}
\label{fig:4}
\end{figure}

The deviations become clearer once more when we consider the intensity ratio of the effective-rate model and the exact
theory, as plotted in the lower part of Fig. \ref{fig:3}.
Remarkably, since the effectice trapping rate $K_{eff}$ is lower than both the reaction-trapping rate 
$\sigma_t C_t$ and the diffusion-trapping rate $K_{diff}$,
the intensity $I_t$ deduced from the effective trapping model is smaller than the exact value.
Deviations from the full model occur throughout the entire intensity regime, although these 
deviations  are less pronounced compared to the two limiting cases 
(fully reaction- or diffusion limited, upper part of Fig. \ref{fig:3}). 
For applications in the analysis of experimental data, 
the accuracy of the effective rate approach [equation~(\ref{eq:K_eff})] can be assessed 
by plotting the intensity ratio (lower part of Fig. \ref{fig:3}) for the respective parameter set.
Irrespectively whether deviations of the effective-rate approach are strong or  minor only, 
the present model founded on diffusion-reaction theory is that which covers the underlying physics
most accurately.

\subsubsection{\label{sec:competitive} Competitive trapping at point defects}
\begin{figure}[ht]
\includegraphics[width=8.5cm]{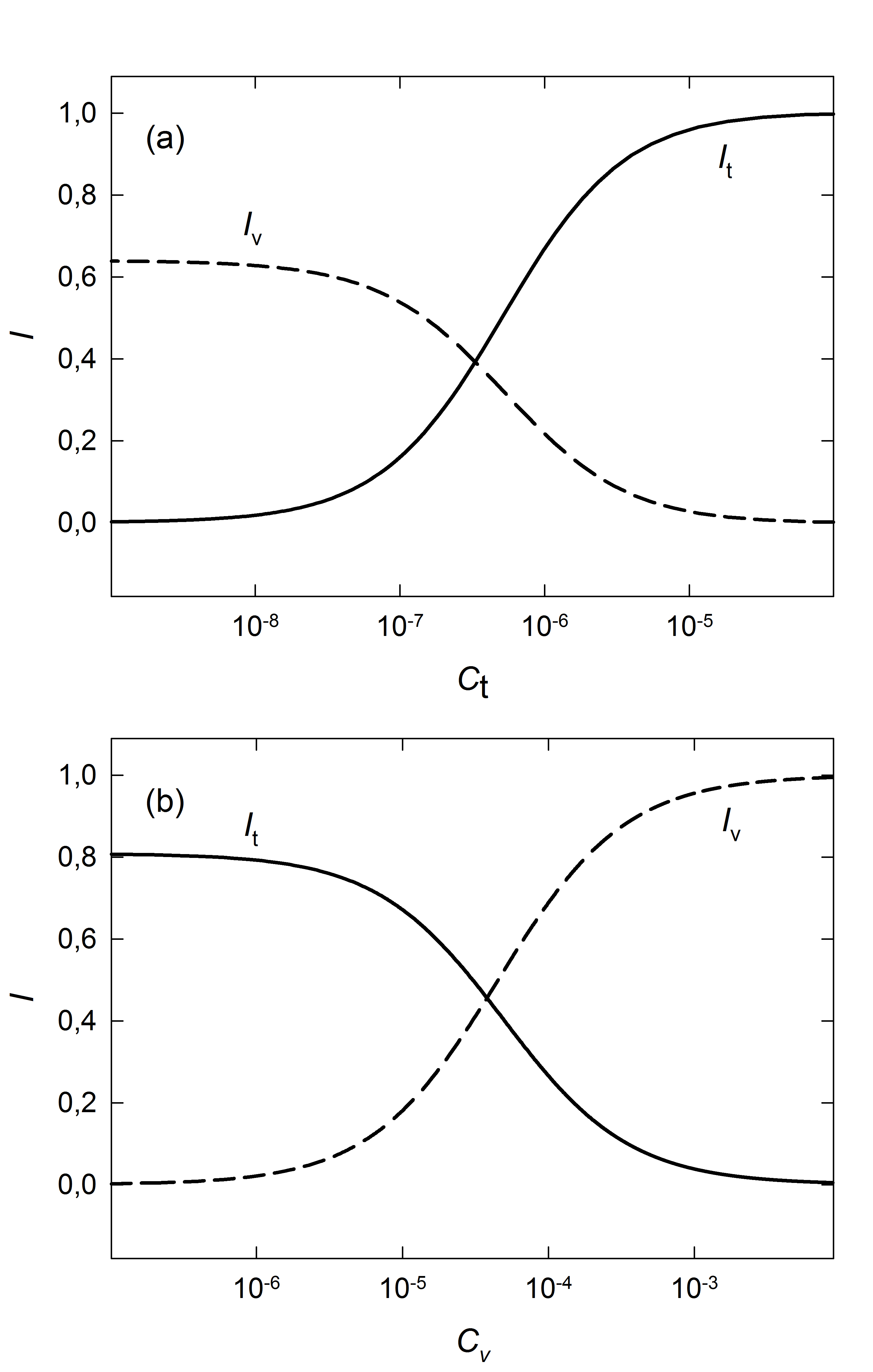}
\caption{Relative intensities
$I_t$ (---) [Eq. (\ref{eq:I_t_general})] of void component $\tau_t$ and $I_v$ ($---$)
[Eq. (\ref{eq:I_v})] of vacancy component $\tau_v$ 
in dependence of (a) void concentration $C_t$ 
and (b) vacancy concentration $C_v$.
Parameters: 
$\tau_v=250$~ps, $\sigma_v=4\times 10^{14}$~s$^{-1}$. $\Omega^{-1} = 6\times 10^{28}$~m$^{-3}$,
others as in Fig. \ref{fig:2}.
(a): $C_v = 10^{-5}$ , (b) $C_t = 10^{-6}$.
}
\label{fig:5}
\end{figure}

Now, we discuss the general case that in addition to diffusion-reaction
controlled trapping at voids also competitive transition-limited trapping
at vacancy-type defects in the lattice occurs (Sect. \ref{sec:vacancies}).
The relative intensities of the void component $I_t$  [Eq. (\ref{eq:I_t_general})] and 
of the vacancy component $I_v$ [Eq. (\ref{eq:I_v})] is plotted in figure \ref{fig:5}
in dependence of void concentration $C_t$ (a) and vacancy concentration $C_v$ (b), 
for a given fixed $C_v$ or $C_t$, respectively. For the 
vacancy-type defect a $e^+$ lifetime component $\tau_v=250$~ps and 
a specific trapping rate $\sigma_v=4\times10^{14}$~s$^{-1}$ \cite{Schaefer87} is assumed; the other parameters 
are the same as used above.
The competitive $e^+$ trapping at voids and  vacancy-type defects becomes evident.
For a given vacancy concentration the
intensity $I_t$ of the void increases and the
intensity $I_v$ of the vacancy  component decreases with increasing 
void concentration due to the increasing fraction of e$^+$ that reaches the
voids (Fig.~\ref{fig:5}.a). Likewise, for a given void concentration,
$I_v$  increases and  $I_t$ decreases with increasing vacancy concentration
(Fig. \ref{fig:5}.b).


\subsubsection{\label{sec:gb} Comparison with $e^+$ trapping at grain boundaries}
\begin{figure}[ht]
\includegraphics[width=8.5cm]{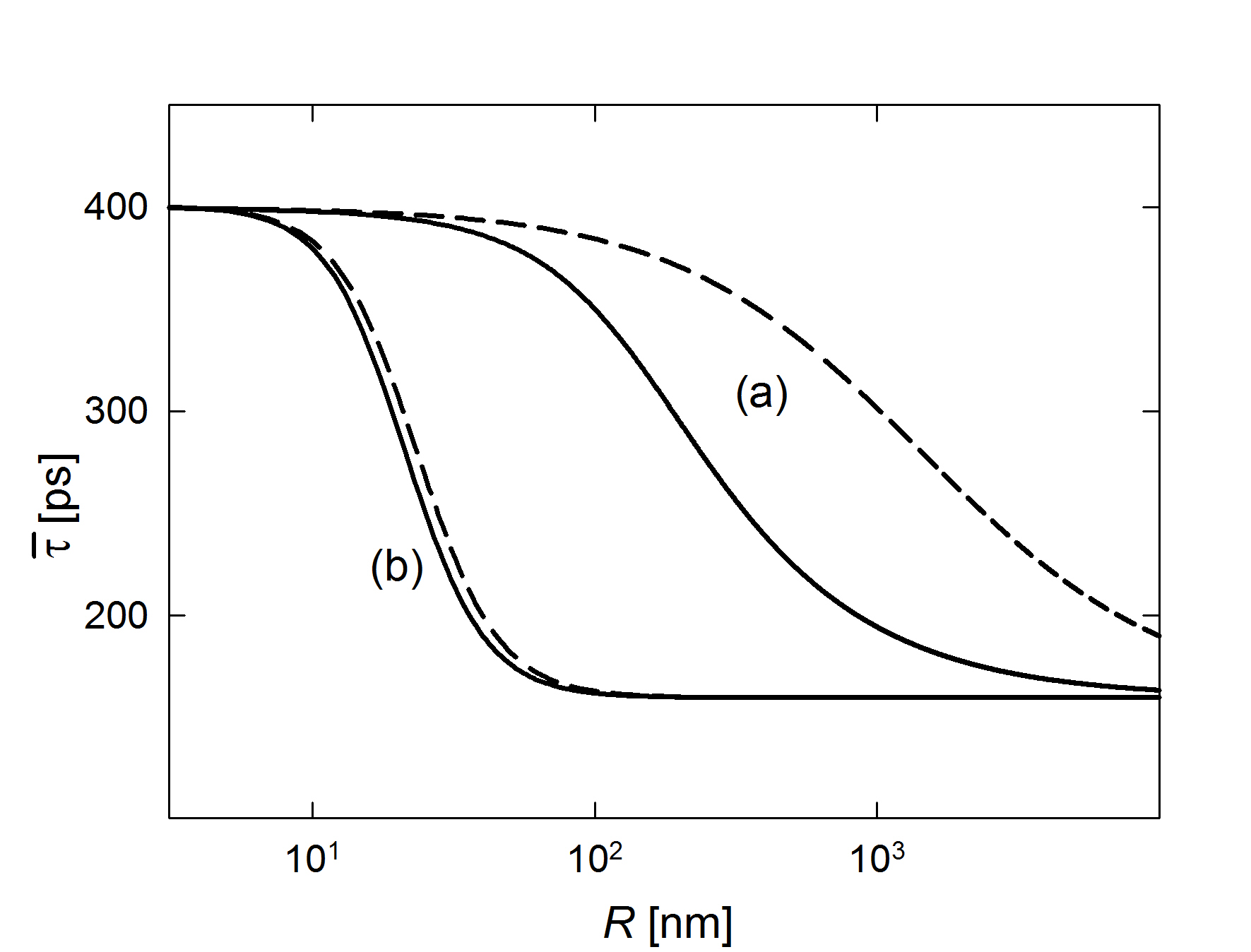}
\caption{Comparison of mean $e^+$ lifetime $\overline{\tau}$ for  diffusion-reaction limited $e^+$ trapping
(a) at grain boundaries of spherical crystallites with radius $R$ \cite{Wuerschum96, Oberdorfer09} and (b)
at extended spherical defects (present work, Fig. \ref{fig:2}).
(a) Exact solution according to eq. (\ref{eq:tauq_GB} \cite{Wuerschum96} (---) and 
solution for standard rate theory ($---$) [Eq. \ref{eq:tauq_rate}] with trapping rate $K = 3 \alpha /R$;
(b) exact solution [Eq. (\ref{eq:tauq})]  (---) and solution
for standard rate theory ($---$) [Eq. (\ref{eq:tauq_rate})] with trapping rate 
$3 \alpha r_0^2 / (R^3-r_0^3)$ [Eq. (\ref{eq:K})].
Parameters as in Fig. \ref{fig:2}.
}
\label{fig:6}
\end{figure}

In the end of this subsection (\ref{sec:voids}), the results of the present model on diffusion-reaction limited $e^+$ trapping
at extended spherical defects will briefly be compared with the corresponding model 
of $e^+$ trapping at grain boundaries of spherical crystallites with radius $R$ \cite{Wuerschum96, Oberdorfer09}.
Whereas in the latter case the surface of the diffusion sphere  
with area $4 \pi R^2$ acts as $e^+$ trap, in the present case with voids of radius $r_0$, the trapping active area 
$4 \pi r_0^2$ is much smaller. Moreover,  the trapping rate $3 \alpha /R$ for grain boundary trapping \cite{Oberdorfer09}
decreases much more slowly with increasing $R$ compared to the trapping rate $3 \alpha r_0^2 / (R^3-r_0^3)$ of 
spherical extended defects with radius $r_0$ [Eq. \ref{eq:K}].
This is the reason why diffusion limitation 
affects the kinetics of $e^+$ trapping at grain boundaries more strongly than in the case of voids
which is nicely  demonstrated in Fig. \ref{fig:6} where the exact solutions are compared with those of infinite diffusivities.
In Fig. \ref{fig:6} the mean $e^+$ lifetime according to the exact solutions  and those of the standard rate theory  
for the two types of extended traps are plotted.
The exact solution for $e^+$ trapping at grain boundaries of spherical crystallites with radius $R$ reads 
\cite{Wuerschum96, Oberdorfer09}
\begin{equation}
\label{eq:tauq_GB}
\overline{\tau} = \tau_f +  (\tau_t-\tau_f) \times 
\frac{3 \alpha L(\gamma_0 R)}{R \gamma_0
\Bigl\{ \alpha + \gamma_0 D L(\gamma_0 R) \Bigl\}} \text{ with } L (z) = \coth z - \frac{1}{z}\, .
\end{equation}
The more stronger deviation between the exact solution and the rate theory in the case of grain boundary trapping is obvious (Fig. \ref{fig:6}).

\subsection{\label{sec:composite}
Larger precipitates: $e^+$-trapping from both sides of precipitate$-$matrix interface }

In Sect.~(\ref{sec:extended}) we extended the model for applying it to larger precipitates taking into account free $e^+$ annihilation
within the precipitate. The $e^+$ trapping from the precipitate into the precipitate$-$matrix interface is handled 
either by rate theory, for special cases where the precipitate radius is well below the $e^+$ diffusion length, or else 
by diffusion-reaction theory, for the more general case that the precipitate radius is in the range of or larger than 
the $e^+$ diffusion length.
With this extension the present model is applicable to a wide variety of structurally complex scenarios, namely to all type
of composite structures where spherical precipitates are embedded in a matrix irrespective of the size and the number density of 
the precipitates.

\begin{figure}[ht]
\includegraphics[width=8.5cm]{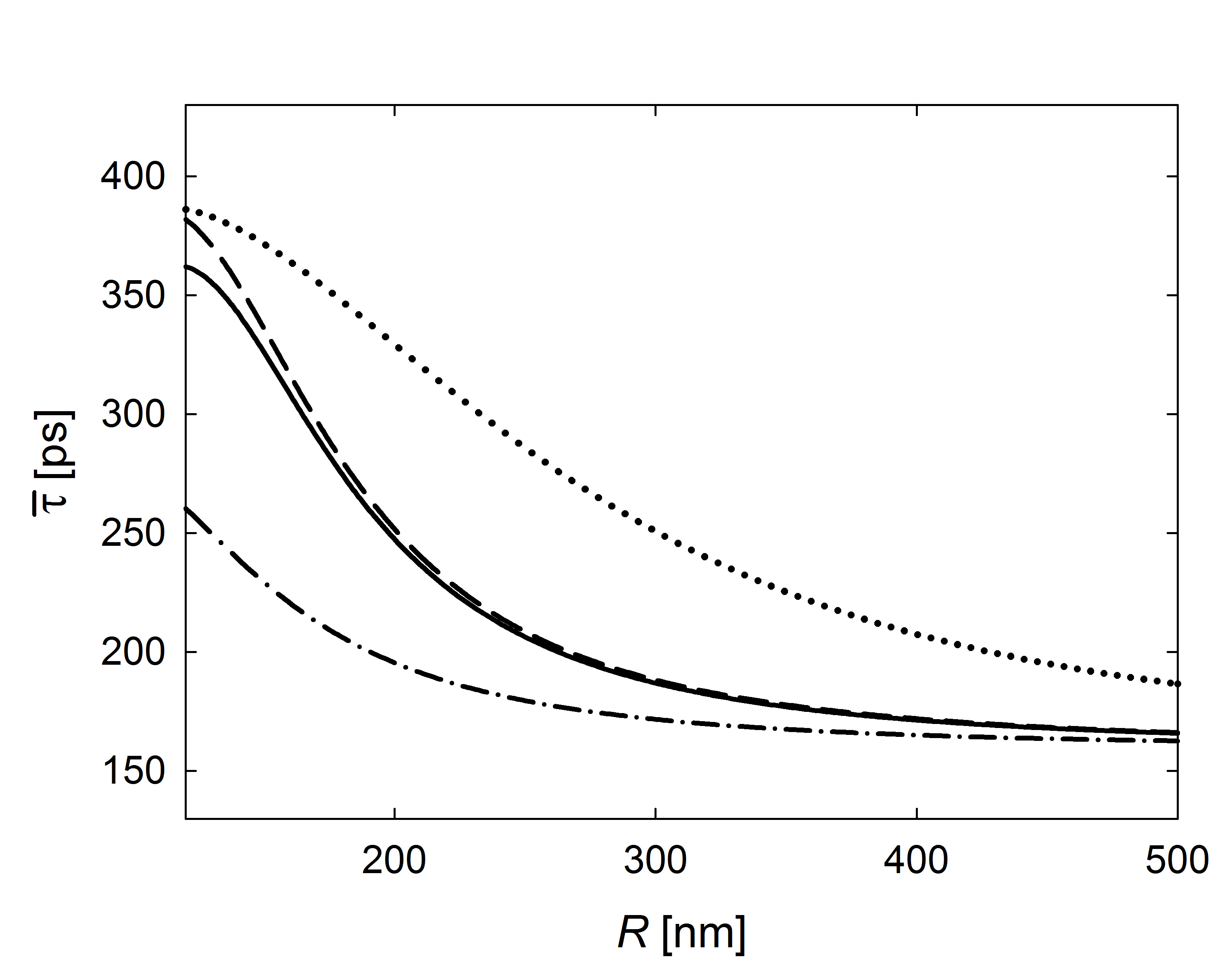}
\caption{Mean $e^+$ lifetime in dependence of radius $R$ 
 (i) for diffusion- and reaction limited $e^+$ trapping into the 
precipitate$-$matrix interface from both the matrix and the precipitate (---) 
[Eq. (\ref{eq:tauq_double_extended})], 
(ii) for diffusion- and reaction limited $e^+$ trapping from the matrix and entirely reaction-controlled 
trapping from the precipitate  ($---$) [Eq. (\ref{eq:tauq_extended})],  (iii)
for entirely reaction-controlled trapping from both the matrix and the precipitate 
($\cdot \cdot \cdot$) [Eq. (\ref{eq:tauq_extended_rate})], and (iv) for rate-model as for (iii), but with 
effective diffusion-limited trapping rate [Eq.~(\ref{eq:K_eff})] ($- \cdot - \cdot -$).
Precipitate radius $r_0 = 100$~nm.  
Other parameters: 
$\tau_f=\tau_c=160$~ps, $\tau_t=400$~ps,
$D=2 \times 10^{-5}$~m$^2$s$^{-1}$, $\alpha=\beta=3 \times 10^3$~ms$^{-1}$.
Note that $R$ is related to the precipitate concentration [Eq. \ref{eq:C_t}].
}
\label{fig:7}
\end{figure}

Whereas for extended defects with smaller size, which were discussed in Sect.~(\ref{sec:voids}),
the deviations between the exact model and the rate theory may be of less relevance since the 
 trapping active area $4 \pi r_0^2$ is small,  
for larger precipitates the diffusion-limitation in any case gets relevant owing
to the much larger trapping active area, similar as for $e^+$ trapping at GBs 
(see Fig.~\ref{fig:6}).
This is demonstrated in Fig.~\ref{fig:7}, where the variation of the mean $e^+$ lifetime with radius $R$
is compared for four different solutions, namely diffusion-limitation of trapping into the 
precipitate$-$matrix interface from both the matrix and the precipitate, from the matrix only, and for
entirely reaction-limited trapping from both sides with standard-trapping rate or with effective diffusion-limited trapping rate.
The latter is obtained by replacing in equation~(\ref{eq:tauq_extended_rate}) the standard-trapping rates by 
the  effective diffusion-limited trapping rate according to equation~(\ref{eq:K_eff}), i.e., 
$3 \alpha r_0^2 (R^3-r_0^3)^{-1}$ by $3 \alpha D r_0^2 R^{-3} (\alpha r_0 +D)^{-1}$ and 
$3 \beta r_0^{-1}$ by $3 \beta D r_0^{-1} (\beta r_0 +D)^{-1}$.

In contrast to the case of small extended defects (Fig.~\ref{fig:6}), for larger precipitates 
(example $r_0=100$~nm) substantial deviations between the solutions occur for the entire concentration regime  
if the diffusion-limitation is neglected (Fig.~\ref{fig:7}). 
Even the rate approach with effective diffusion-limited trapping rate,
which at least for small extended defects is a reasonable approximation (Sect.~\ref{sec:effective_rate}, Fig.~\ref{fig:4}),
turns out to be completely inadequate for the larger precipitate size.
The deviations are much less if the diffusion-limitation is only neglected
for the trapping from the precipitate into the interface, since the precipitate size (in contrast to the precipitate distance) remains in the range of the $e^+$ diffusion length independent of the 
precipitate concentration. Anyhow, for a precise description even for such small precipitate sizes, 
the exact theory of diffusion- and reaction controlled trapping has to be applied for
the trapping from the interior of the precipitates.

Finally, we compare this model with that presented by Dryzek \cite{dryzek1999, dryzek2016}
for studying recrystallization in highly deformed metals. In that case recrystallized grains are embedded in a highly deformed matrix. Diffusion-limited $e^+$ trapping occurs from the grains into the matrix,
whereas within the matrix saturation trapping of $e^+$ prevails due to the high defect density.
In this sense, the model of Dryzek represents an extension of the diffusion-reaction theory for trapping at grain boundaries, where instead of GBs a surrounding deformed matrix is considered. The model presented here, represents a further extension where  diffusion- and reaction-controlled trapping also from the matrix into the interfaces is considered.

\section{\label{conclusion} Conclusion}

The present model with the exact solution
of  the diffusion-reaction theory for the $e^+$ trapping at 
 extended spherical defects  
and competitive transition-limited trapping at  atomic defects 
yields a basis for the quantitative description of the
 $e^+$  behaviour in materials with complex defect structure.
It could be shown that the model includes  as special cases
the simple trapping model and the entirely diffusion-limited trapping,
but both of these limiting cases represent approximations, only.
For the full model, closed-form expressions were obtained for
the mean positron lifetime  $\overline{\tau}$ and for the intensities
of the e$^+$ lifetime components associated with
trapping. This exact model allowed a quantitative assessment of  the usual approach, 
which takes  diffusion limitation for the trapping at voids into account 
by effective diffusion-trapping rates. The present closed-form solutions also  
renders this effective rate approach unnecessary.

The presented theory  goes even much  far beyond existing models,  
since it is not only applicable to small extended defects (such as voids or clusters), but also 
to larger precipitates where positron trapping from the precipitates into the
precipitate$-$matrix interface is taken into consideration.
Therefore, the model presents the basis for studying all type
of composite structures where spherical precipitates are embedded in a matrix irrespective of their size and  their number density.

\begin{acknowledgments}
The senior author (R.W.) dedicates this work to Alfred Seeger 
whose numerous pioneering works also included modeling of positron annihilation.
This work was performed in the framework of the inter-university cooperation of TU Graz and Uni Graz on natural science (NAWI
Graz).
\end{acknowledgments}
\newpage

\begin{thebibliography}{24}
\expandafter\ifx\csname natexlab\endcsname\relax\def\natexlab#1{#1}\fi
\expandafter\ifx\csname bibnamefont\endcsname\relax
  \def\bibnamefont#1{#1}\fi
\expandafter\ifx\csname bibfnamefont\endcsname\relax
  \def\bibfnamefont#1{#1}\fi
\expandafter\ifx\csname citenamefont\endcsname\relax
  \def\citenamefont#1{#1}\fi
\expandafter\ifx\csname url\endcsname\relax
  \def\url#1{\texttt{#1}}\fi
\expandafter\ifx\csname urlprefix\endcsname\relax\def\urlprefix{URL }\fi
\providecommand{\bibinfo}[2]{#2}
\providecommand{\eprint}[2][]{\url{#2}}

\bibitem[{\citenamefont{Hautoj\"{a}rvi}(1979)}]{Hautojaervi79}
\bibinfo{author}{\bibfnamefont{P.}~\bibnamefont{Hautoj\"{a}rvi}},
  \emph{\bibinfo{title}{Positrons in Solids}} (\bibinfo{publisher}{Springer},
  \bibinfo{address}{Berlin}, \bibinfo{year}{1979}).

\bibitem[{\citenamefont{Keeble et~al.}(2012)\citenamefont{Keeble, Brossmann,
  Puff, and W\"{u}rschum}}]{Brossmann12}
\bibinfo{author}{\bibfnamefont{D.}~\bibnamefont{Keeble}},
  \bibinfo{author}{\bibfnamefont{U.}~\bibnamefont{Brossmann}},
  \bibinfo{author}{\bibfnamefont{W.}~\bibnamefont{Puff}}, \bibnamefont{and}
  \bibinfo{author}{\bibfnamefont{R.}~\bibnamefont{W\"{u}rschum}}, in
  \emph{\bibinfo{booktitle}{Characterization of Materials}}, edited by
  \bibinfo{editor}{\bibfnamefont{E.}~\bibnamefont{Kaufmann}}
  (\bibinfo{publisher}{John Wiley \& Sons}, \bibinfo{year}{2012}), p.
  \bibinfo{pages}{1899}.

\bibitem[{\citenamefont{Krause-Rehberg and Leipner}(1999)}]{Krause-Rehberg99}
\bibinfo{author}{\bibfnamefont{R.}~\bibnamefont{Krause-Rehberg}}
  \bibnamefont{and} \bibinfo{author}{\bibfnamefont{H.}~\bibnamefont{Leipner}},
  \emph{\bibinfo{title}{Positron Annihilation in Semiconductors}}
  (\bibinfo{publisher}{Springer}, \bibinfo{address}{Berlin},
  \bibinfo{year}{1999}).

\bibitem[{\citenamefont{Puska and Nieminen}(1994)}]{Puska94}
\bibinfo{author}{\bibfnamefont{M.~J.} \bibnamefont{Puska}} \bibnamefont{and}
  \bibinfo{author}{\bibfnamefont{R.~M.} \bibnamefont{Nieminen}},
  \bibinfo{journal}{Rev. Mod. Phys.} \textbf{\bibinfo{volume}{66}},
  \bibinfo{pages}{841} (\bibinfo{year}{1994}).

\bibitem[{\citenamefont{Brandt and Paulin}(1972)}]{Brandt72}
\bibinfo{author}{\bibfnamefont{W.}~\bibnamefont{Brandt}} \bibnamefont{and}
  \bibinfo{author}{\bibfnamefont{R.}~\bibnamefont{Paulin}},
  \bibinfo{journal}{Phys. Rev. B} \textbf{\bibinfo{volume}{5}},
  \bibinfo{pages}{2430} (\bibinfo{year}{1972}).

\bibitem[{\citenamefont{Dupasquier et~al.}(1993)\citenamefont{Dupasquier,
  Romero, and Somoza}}]{Dupasquier93}
\bibinfo{author}{\bibfnamefont{A.}~\bibnamefont{Dupasquier}},
  \bibinfo{author}{\bibfnamefont{R.}~\bibnamefont{Romero}}, \bibnamefont{and}
  \bibinfo{author}{\bibfnamefont{A.}~\bibnamefont{Somoza}},
  \bibinfo{journal}{Phys. Rev. B} \textbf{\bibinfo{volume}{48}},
  \bibinfo{pages}{9235} (\bibinfo{year}{1993}).

\bibitem[{\citenamefont{W\"urschum and Seeger}(1996)}]{Wuerschum96}
\bibinfo{author}{\bibfnamefont{R.}~\bibnamefont{W\"urschum}} \bibnamefont{and}
  \bibinfo{author}{\bibfnamefont{A.}~\bibnamefont{Seeger}},
  \bibinfo{journal}{Phil. Mag. A} \textbf{\bibinfo{volume}{73}},
  \bibinfo{pages}{1489} (\bibinfo{year}{1996}).

\bibitem[{\citenamefont{K\"{o}gel}(1996)}]{Koegel96}
\bibinfo{author}{\bibfnamefont{G.}~\bibnamefont{K\"{o}gel}},
  \bibinfo{journal}{Appl. Phys. A} \textbf{\bibinfo{volume}{63}},
  \bibinfo{pages}{227} (\bibinfo{year}{1996}).

\bibitem[{\citenamefont{Dryzek}(1999)}]{dryzek1999}
\bibinfo{author}{\bibfnamefont{J.}~\bibnamefont{Dryzek}},
  \bibinfo{journal}{Acta Physica Polonica A} \textbf{\bibinfo{volume}{95}},
  \bibinfo{pages}{539} (\bibinfo{year}{1999}).

\bibitem[{\citenamefont{\u{C}i\u{z}ek et~al.}(2002)\citenamefont{\u{C}i\u{z}ek,
  Proch\'{a}zka, Cieslar, Ku\u{z}el, Kuriplach, Chmelik, Stul\'{i}kov\'{a},
  Be\u{c}v\'{a}\u{r}, Melikhova, and Islamgaliev}}]{Cizek02}
\bibinfo{author}{\bibfnamefont{J.}~\bibnamefont{\u{C}i\u{z}ek}},
  \bibinfo{author}{\bibfnamefont{I.}~\bibnamefont{Proch\'{a}zka}},
  \bibinfo{author}{\bibfnamefont{M.}~\bibnamefont{Cieslar}},
  \bibinfo{author}{\bibfnamefont{R.}~\bibnamefont{Ku\u{z}el}},
  \bibinfo{author}{\bibfnamefont{J.}~\bibnamefont{Kuriplach}},
  \bibinfo{author}{\bibfnamefont{F.}~\bibnamefont{Chmelik}},
  \bibinfo{author}{\bibfnamefont{I.}~\bibnamefont{Stul\'{i}kov\'{a}}},
  \bibinfo{author}{\bibfnamefont{F.}~\bibnamefont{Be\u{c}v\'{a}\u{r}}},
  \bibinfo{author}{\bibfnamefont{O.}~\bibnamefont{Melikhova}},
  \bibnamefont{and}
  \bibinfo{author}{\bibfnamefont{R.}~\bibnamefont{Islamgaliev}},
  \bibinfo{journal}{Phys. Rev. B.} \textbf{\bibinfo{volume}{65}},
  \bibinfo{pages}{094106} (\bibinfo{year}{2002}).

\bibitem[{\citenamefont{Oberdorfer and W\"urschum}(2009)}]{Oberdorfer09}
\bibinfo{author}{\bibfnamefont{B.}~\bibnamefont{Oberdorfer}} \bibnamefont{and}
  \bibinfo{author}{\bibfnamefont{R.}~\bibnamefont{W\"urschum}},
  \bibinfo{journal}{Phys. Rev. B.} \textbf{\bibinfo{volume}{79}},
  \bibinfo{pages}{184103} (\bibinfo{year}{2009}).

\bibitem[{\citenamefont{Dryzek et~al.}(1998)\citenamefont{Dryzek, Czapla, and
  Kusior}}]{dryzek1998}
\bibinfo{author}{\bibfnamefont{J.}~\bibnamefont{Dryzek}},
  \bibinfo{author}{\bibfnamefont{A.}~\bibnamefont{Czapla}}, \bibnamefont{and}
  \bibinfo{author}{\bibfnamefont{E.}~\bibnamefont{Kusior}},
  \bibinfo{journal}{J. Phys. Condens. Matter} \textbf{\bibinfo{volume}{10}},
  \bibinfo{pages}{10827} (\bibinfo{year}{1998}).

\bibitem[{\citenamefont{Nieminen and Laakkonen}(1979)}]{Nieminen79a}
\bibinfo{author}{\bibfnamefont{R.}~\bibnamefont{Nieminen}} \bibnamefont{and}
  \bibinfo{author}{\bibfnamefont{J.}~\bibnamefont{Laakkonen}},
  \bibinfo{journal}{Appl. Phys.} \textbf{\bibinfo{volume}{20}},
  \bibinfo{pages}{181} (\bibinfo{year}{1979}).

\bibitem[{\citenamefont{Bentzon and Evans}(1990)}]{Bentzon90}
\bibinfo{author}{\bibfnamefont{M.}~\bibnamefont{Bentzon}} \bibnamefont{and}
  \bibinfo{author}{\bibfnamefont{J.}~\bibnamefont{Evans}}, \bibinfo{journal}{J.
  Phys. Condens. Matter} \textbf{\bibinfo{volume}{2}}, \bibinfo{pages}{10165}
  (\bibinfo{year}{1990}).

\bibitem[{\citenamefont{Eldrup and Singh}(2003)}]{Eldrup03}
\bibinfo{author}{\bibfnamefont{M.}~\bibnamefont{Eldrup}} \bibnamefont{and}
  \bibinfo{author}{\bibfnamefont{B.}~\bibnamefont{Singh}}, \bibinfo{journal}{J.
  Nucl. Mater.} \textbf{\bibinfo{volume}{323}}, \bibinfo{pages}{346}
  (\bibinfo{year}{2003}).

\bibitem[{\citenamefont{Dryzek}(1998)}]{dryzek1998diffusion}
\bibinfo{author}{\bibfnamefont{J.}~\bibnamefont{Dryzek}},
  \bibinfo{journal}{phys. stat. sol. (b)} \textbf{\bibinfo{volume}{209}},
  \bibinfo{pages}{3} (\bibinfo{year}{1998}).

\bibitem[{\citenamefont{Nieminen et~al.}(1979)\citenamefont{Nieminen,
  Laakkonen, Hautoj\"arvi, and Vehanen}}]{Nieminen79}
\bibinfo{author}{\bibfnamefont{R.}~\bibnamefont{Nieminen}},
  \bibinfo{author}{\bibfnamefont{J.}~\bibnamefont{Laakkonen}},
  \bibinfo{author}{\bibfnamefont{P.}~\bibnamefont{Hautoj\"arvi}},
  \bibnamefont{and} \bibinfo{author}{\bibfnamefont{A.}~\bibnamefont{Vehanen}},
  \bibinfo{journal}{Phys. Rev. B} \textbf{\bibinfo{volume}{19}},
  \bibinfo{pages}{1397} (\bibinfo{year}{1979}).

\bibitem[{\citenamefont{W\"urschum and Seeger}(1995)}]{Wuerschum95}
\bibinfo{author}{\bibfnamefont{R.}~\bibnamefont{W\"urschum}} \bibnamefont{and}
  \bibinfo{author}{\bibfnamefont{A.}~\bibnamefont{Seeger}},
  \bibinfo{journal}{Zeitschr. Phys. Chemie} \textbf{\bibinfo{volume}{192}},
  \bibinfo{pages}{47} (\bibinfo{year}{1995}).

\bibitem[{\citenamefont{Olver}(2010)}]{olver2010nist}
\bibinfo{author}{\bibfnamefont{F.~W.} \bibnamefont{Olver}},
  \emph{\bibinfo{title}{NIST Handbook of Mathematical Functions}}
  (\bibinfo{publisher}{Cambridge University Press}, \bibinfo{year}{2010}).

\bibitem[{\citenamefont{Dryzek}(2002)}]{dryzek2002}
\bibinfo{author}{\bibfnamefont{J.}~\bibnamefont{Dryzek}},
  \bibinfo{journal}{phys. stat. sol. (b)} \textbf{\bibinfo{volume}{229}},
  \bibinfo{pages}{1163} (\bibinfo{year}{2002}).

\bibitem[{\citenamefont{Nambissan et~al.}(1989)\citenamefont{Nambissan, Sen,
  and Viswanathan}}]{Nambissan1989}
\bibinfo{author}{\bibfnamefont{P.}~\bibnamefont{Nambissan}},
  \bibinfo{author}{\bibfnamefont{P.}~\bibnamefont{Sen}}, \bibnamefont{and}
  \bibinfo{author}{\bibfnamefont{B.}~\bibnamefont{Viswanathan}}, in
  \emph{\bibinfo{booktitle}{Positron Annihilation}}, edited by
  \bibinfo{editor}{\bibnamefont{L.Dorikens-Vanpraet}},
  \bibinfo{editor}{\bibnamefont{M.Dorikens}}, \bibnamefont{and}
  \bibinfo{editor}{\bibnamefont{D.Segers}} (\bibinfo{publisher}{World
  Scientific Publ.}, \bibinfo{year}{1989}), p. \bibinfo{pages}{434}.

\bibitem[{\citenamefont{Frank and Seeger}(1974)}]{Seeger74}
\bibinfo{author}{\bibfnamefont{W.}~\bibnamefont{Frank}} \bibnamefont{and}
  \bibinfo{author}{\bibfnamefont{A.}~\bibnamefont{Seeger}},
  \bibinfo{journal}{Appl. Phys.} \textbf{\bibinfo{volume}{3}},
  \bibinfo{pages}{61} (\bibinfo{year}{1974}).

\bibitem[{\citenamefont{Schaefer}(1987)}]{Schaefer87}
\bibinfo{author}{\bibfnamefont{H.-E.} \bibnamefont{Schaefer}},
  \bibinfo{journal}{phys. stat. sol. (a)} \textbf{\bibinfo{volume}{103}},
  \bibinfo{pages}{97} (\bibinfo{year}{1987}).

\bibitem[{\citenamefont{Dryzek et~al.}(2016)\citenamefont{Dryzek, Wrobel, and
  Dryzek}}]{dryzek2016}
\bibinfo{author}{\bibfnamefont{J.}~\bibnamefont{Dryzek}},
  \bibinfo{author}{\bibfnamefont{M.}~\bibnamefont{Wrobel}}, \bibnamefont{and}
  \bibinfo{author}{\bibfnamefont{E.}~\bibnamefont{Dryzek}},
  \bibinfo{journal}{phys. stat. sol. (b)} \textbf{\bibinfo{volume}{253}},
  \bibinfo{pages}{2031} (\bibinfo{year}{2016}).

\end{thebibliography}


\newpage
\newpage

\newpage

\newpage

\newpage

\newpage

\newpage

\newpage

\begin{center}
{\bf {\Large  Supplement to Physical Review B 97 (2018) 224108}}\\
Title: Diffusion-reaction model for positron trapping and annihilation at spherical extended  defects
and in precipitate$-$matrix composites \\
DOI: 10.1103/PhysRevB.97.224108 \\[.3cm]
Roland W\"{u}rschum, Laura Resch, and Gregor Klinser \\
{\it Institute of Materials Physics, Graz University of Technology, Petersgasse 16, A-8010 Graz, Austria 
(email: wuerschum@tugraz.at)} \\
April 11, 2019 \\[.3cm]
{\bf Supplement to Section III, A}
\end{center} 
{\bf [i] Intensities associated with positron annihilation from the free state\footnote{Intensities $I_{0,j}$
as mentioned but not explicitly quoted in Section III, A (see text in [1] after eq. (34)).
}} \\
The sequence of intensities for poles $p = - \lambda_{0,j}$ given by Eq.~(33) of [1] read: 
\begin{eqnarray}
\label{eq:I_j}
I_{0,j} = \frac{K  (\tau_f^{-1}-\tau_t^{-1})2 (1 + \gamma^{\star 2}_j R^2)}{(\lambda_{0,j} - \tau_f^{-1})(\lambda_{0,j} -\tau_t^{-1})} \times \\ \nonumber
\Bigl\{ \frac{D \gamma^{\star 2}_j}{\alpha r_0^2} \Bigl[R^3-r_0^3 + (R-r_0)r_0^2R^2(\gamma^{\star 2}_j+\frac{\alpha^2}{D^2})+
\frac{\alpha r_0}{D}R^2(2R-r_0) \Bigr]  -1 - \frac{\alpha r_0}{D}\Bigr\}^{-1}
\end{eqnarray}
with 
\begin{equation}
\label{eq:gamma_i}
\gamma^{\star 2}_j = \frac{\lambda_{0,j} - \tau_{f}^{-1}}{D} 
\end{equation}
and 
\begin{equation}
K= \frac{3 \alpha r_0^2}{R^3-r_0^3} \,.
\end{equation}
As usual for this kind of diffusion-reaction problems, the intensities of these decay rates rapidly decrease (see eaxmple, Table I).
\\[.2cm]
\underline{Limiting case of entirely reaction-controlled trapping}
\\
Expansion of $\tan(z)$  in Eq.~(33) [1] up to third order yields for $D \rightarrow \infty$:
\begin{equation}
\lambda_0 = \frac{1}{\tau_0} = \frac{1}{\tau_f} + K \, .
\label{lambda_0}
\end{equation}
Inserting $\gamma^{\star 2} = K/D$ in Eq. (\ref{eq:I_j}) yields 
with $\lambda_{0} - \tau_f^{-1} = K$ for $D \rightarrow \infty$:
\begin{equation}
I_{0} \simeq \frac{K  (\tau_f^{-1}-\tau_t^{-1})2}{(\lambda_{0} - \tau_f^{-1})(\lambda_{0} -\tau_t^{-1})
\Bigl\{\frac{K}{\alpha r_0^2} (R^3-r_0^3)-1 \Bigr\}} = \frac{\tau_f^{-1}-\tau_t^{-1}}{\tau_0^{-1} - \tau_f^{-1}} \, .
\label{I_0}
\end{equation}
The equations (\ref{lambda_0}) and (\ref{I_0}) correspond to the solutions of the standard two-state trapping theory, i.e., 
the present model includes in the limiting case of high positron diffusivity  the solution of the standard-rate theory. \\[1cm]
{\bf [ii] Inspection\footnote{Detailed documentation of statement in Section III, A (see second paragraph after eq. (34) [1]).} of pole $p = - 1/\tau_f$} \\
Expansion of $\tanh(z)$ up to third order yields for the second fraction $F$ of Eq.~(29) [1]: 
\begin{equation}
F = \frac{1/3 \gamma^3 (R^3 - r_0^3)}{1/3 \gamma^3 (R^3 - r_0^3) + \frac{\alpha r_0}{D} (\gamma r_0 + 1/3\gamma^3 \hat{R}^3)} = 
\frac{\gamma^2}{\gamma^2 + \frac{K}{D}(1 + \gamma^2\frac{\hat{R}^3}{3r_0})}
\label{eq:F}
\end{equation}
By means of eq. (\ref{eq:F})  $\tilde{n}(p)$ (Eq.~(29) of [1]) can be written as:
\begin{equation}
\tilde{n}(p) = \frac{\tau_t^{-1}+p+K + \frac{K}{D} \frac{\hat{R}^3}{3r_0} (\tau_t^{-1}+p)}{(\tau_t^{-1}+p)\{\tau_f^{-1}+p + K [1+ \frac{\hat{R}^3}{3Dr_0} (\tau_f^{-1}+p)]\}}
\label{eq:n(p)}
\end{equation}
\underline{Limiting cases}
\begin{itemize}
\item
For the limiting case $p = - 1/\tau_f$ eq. (\ref{eq:n(p)}) reads:
\begin{equation}
\tilde{n}( -\tau_f^{-1}) = \frac{\tau_t^{-1}-\tau_f^{-1}+K +  \frac{K}{D} \frac{\hat{R}^3}{3r_0} (\tau_t^{-1}-\tau_f^{-1})}{(\tau_t^{-1}-\tau_f^{-1})K} \, .
\label{eq:limit}
\end{equation}
Eq. (\ref{eq:limit}) shows there is no singularity for $p = - 1/\tau_f$. Since the residue of a removable singularity is zero, 
the positron lifetime spectrum does not contain a component $\tau_f$, as expected in the case of positron trapping.
\item
For the limiting case $D \rightarrow \infty$ eq. (\ref{eq:n(p)}) reads:
\begin{equation}
\tilde{n}(p) = \frac{\tau_t^{-1}+p+K}{(\tau_t^{-1}+p)(\tau_f^{-1}+p + K)} \, .
\label{eq:n(p)_limit}
\end{equation}
Eq. (\ref{eq:n(p)_limit}) yields two poles which correspond to 
the two positron lifetime components $\tau_t$ and $1/(\tau_f^{-1} + K)$ according 
to the the standard two-state trapping model, i.e., 
the present model includes in the limiting case of high positron diffusivity  the solution of the standard-rate theory.
\end{itemize} 

\begin{table}
\centering
\caption{Sequence of first-order poles $p=-\lambda_{0,j}$ (eq. (33), [1]) and their corresponding intensities $I_{0,j}$ (eq. (\ref{eq:I_j})), intensity $I_t$  (eq. (30), [1]) of the positron lifetime component $\tau_t$ and sum of all intensities $I_{total}$ for
different values of $R$.
Parameters (as for Fig.~2 of [1]): $\tau_f=160$~ps, $\tau_t=400$~ps,
$D=2 \times 10^{-5}$~m$^2$s$^{-1}$, $\alpha=3 \times 10^3$~ms$^{-1}$,
$r_0 = 3$~nm.
}
\begin{tabular}{cl||c|c|c|c|}
\multirow{5}{1.5em}{\begin{sideways} \textbf{$R=10$ nm}\end{sideways}}
& $j$ & $\lambda_{0,j}$ [$s^ {-1}$] 	& $I_{0,j}$ [\%]  					&  $I_{t}$ [\%] 					& $I_{total}$ [\%] 				\\\cline{2-6}
& 1 	& 7.375 $\times$ $10^{10}$		& 5.26										&	\multirow{3}{*}{94.74}	&	\multirow{3}{*}{100.00} \\
& 2		& 5.581 $\times$ $10^{12}$		& 7.74 $\times$ $10^{-5}$	&													&													\\
& 3		& 1.805 $\times$ $10^{13}$		& 3.46 $\times$ $10^{-6}$	&													&													\\
\\
\end{tabular}
\begin{tabular}{cl||c|c|c|c|}
\multirow{5}{1.5em}{\begin{sideways} \textbf{$R=25$ nm}\end{sideways}}
& $j$ & $\lambda_{0,j}$ [$s^ {-1}$] 	& $I_{0,j}$ [\%] 					&  $I_{t}$ [\%] 				& $I_{total}$ [\%] 			\\\cline{2-6}
& 1 	& 1.008 $\times$ $10^{10}$		& 49.44										&	\multirow{3}{*}{50.56}	&	\multirow{3}{*}{100.00} \\
& 2		& 7.117 $\times$ $10^{11}$		& 1.91 $\times$ $10^{-4}$	&													&													\\
& 3		& 2.113 $\times$ $10^{12}$		& 1.68 $\times$ $10^{-5}$	&													&													\\
\\
\end{tabular}
\begin{tabular}{cl||c|c|c|c|}
\multirow{5}{1.5em}{\begin{sideways} \textbf{$R=50$ nm}\end{sideways}}
& $j$ & $\lambda_{0,j}$ [$s^ {-1}$] 	& $I_{0,j}$ [\%] 					&  $I_{t}$ [\%] 					& $I_{total}$ [\%]  				\\\cline{2-6}
& 1 	& 6.712 $\times$ $10^{9}$			& 89.01										&	\multirow{4}{*}{10.99}	&	\multirow{4}{*}{100.00} \\
& 2		& 1.745 $\times$ $10^{11}$		& 2.23 $\times$ $10^{-4}$	&													&													\\
& 3		& 5.037 $\times$ $10^{11}$		& 2.35 $\times$ $10^{-5}$	&													&													\\
\end{tabular}
\end{table}

[1] R. W\"{u}rschum, L.  Resch, and G. Klinser, Phys. Rev. B {\bf 97} (2018) 224108.

\end{document}